\documentclass[a4paper,fleqn]{cas-dc}
\usepackage{amsmath} 
\usepackage{soul} 
\usepackage{color, xcolor} 
\usepackage{lineno}
\usepackage[numbers]{natbib}
\usepackage[boxed,commentsnumbered,ruled,linesnumbered]{algorithm2e}
\usepackage{graphicx}

\usepackage{hyperref}
\hypersetup{colorlinks=true,linkcolor=blue,citecolor=blue,urlcolor=black}

\newtheorem{definition}{Definition} 
 
\newtheorem{property}{Property}  
\newtheorem{strategy}{Strategy}

\begin{document}
\shorttitle{Discovering Top-$k$ Periodic and High-Utility Patterns} 
\shortauthors{Q. Zhou \textit{et al.}}

\title [mode = title]{Discovering Top-$k$ Periodic and High-Utility Patterns}   

\author[1]{Qingfeng Zhou}
\ead{qingfeng014@gmail.com}
\address[1]{College of Cyber Security, Jinan University, Guangzhou 510632, China}

\author[1]{Wensheng Gan}
\cortext[cor1]{Corresponding author}
\ead{wsgan001@gmail.com}
\cormark[1]

\author[2]{Guoting Chen}
\ead{guoting.chen@univ-lille.fr}
\address[2]{School of Science, Great Bay University, Dongguan 523000, China}

\begin{abstract}
     With a user-specified minimum utility threshold (\textit{minutil}), periodic high-utility pattern mining (PHUPM) aims to identify high-utility patterns that occur periodically in a transaction database. A pattern is deemed periodic if its period aligns with the periodicity constraint set by the user. However, users may not be interested in all periodic high-utility patterns (PHUPs). Moreover, setting \textit{minutil} in advance is also a challenging issue. To address these issues, our research introduces an algorithm called TPU for extracting the most significant top-$k$ periodic and high-utility patterns that may or may not include negative utility values. This TPU algorithm utilizes positive and negative utility lists (PNUL) and period-estimated utility co-occurrence structure (PEUCS) to store pertinent itemset information. Additionally, it incorporates the periodic real item utility (PIU), periodic co-occurrence utility descending (PCUD), and periodic real utility  (PRU) threshold-raising strategies to elevate the thresholds rapidly. By using the proposed threshold-raising strategies, the runtime was reduced by approximately 5\% on the datasets used in the experiments. Specifically, the runtime was reduced by up to 50\% on the mushroom\_negative and kosarak\_negative datasets, and by up to 10\% on the chess\_negative dataset. Memory consumption was reduced by about 2\%, with the largest reduction of about 30\% observed on the mushroom\_negative dataset. Through extensive experiments, we have demonstrated that our algorithm can accurately and effectively extract the top-$k$ periodic high-utility patterns. This paper successfully addresses the top-$k$ mining issue and contributes to data science. Furthermore, the applications of the proposed algorithm in engineering include data mining, expert systems, and web intelligence in various fields, such as smart retail, cyberspace security, and risk prediction. The code and datasets are publicly available at \href{https://github.com/DSI-Lab1/TPU}{https://github.com/DSI-Lab1/TPU.}
\end{abstract}


\maketitle

\section{Introduction}  \label{sec: introduction}

The rapid development of artificial intelligence has driven advancements in various fields, such as time series forecasting \cite{singh2021fqtsfm,singh2025novel}. However, much of this research tends to focus on final prediction or classification accuracy, rather than directly exploring and uncovering the intrinsic high-value information within the data itself. Pattern mining, e.g., frequent itemset mining (FIM) \cite{jazayeri2024frequent}, is a pivotal task within the domain of data mining \cite{li2024opf,rage2024pami}, aimed at uncovering interesting itemsets from a database. The primary objective of FIM is to identify itemsets in a transaction database with a frequency or support higher than a specified threshold \cite{agrawal1994Apriori,han2000FP-Growth}. However, the FIM algorithms typically focus solely on itemsets within the transactions of a database, disregarding the significance of individual items. In many practical scenarios, frequent itemsets may not necessarily translate to high profits, while less frequent itemsets may yield higher profits. To address the challenges encountered by FIM algorithms, many studies have considered the weights of items within each transaction and have proposed several high-utility itemset mining (HUIM) algorithms \cite{cheng2023efficient,nguyen2023parallel,vlashejerdi2025efficient}. HUIM focuses on finding all itemsets with utility values equal to or greater than a specified minimum utility threshold (\textit{minutil}). Several types of HUIM algorithms exist, including two-phase algorithms such as Two-Phase \cite{liu2005two-PhaseAlgorithm}, tree-based algorithms like UP-Growth \cite{tseng2010UP-Growth}, utility-list-based algorithms such as HUI-Miner \cite{liu2012mining}, data-projection-based algorithms such as EFIM \cite{zida2015efim}, as well as more recent approaches like DPHIM \cite{2023HUIOnMulticoreProcessors}, R-Miner \cite{sra2024residual}, EMHUN \cite{tung2024efficient}, and FOTH \cite{yan2024efficient}. Although there are many efficient HUIM algorithms in practical scenarios, a significant challenge remains the need for users to predefine the \textit{minutil} threshold. Setting a suitable \textit{minutil} is inherently difficult. Accordingly, there is an urgent need for an effective method to address the \textit{minutil} setting problem.

Typically, users are not interested in all the mined HUIs but instead focus on the top-$k$ itemsets ranked by their highest utility. Thus, several studies have proposed several top-$k$ high-utility itemset mining (top-$k$ HUIM) algorithms \cite{han2024mining,liu2023mining,luna2023efficient}. The objective of top-$k$ HUIM algorithms is to discover high-utility itemsets based on the user-specified value of $k$, which denotes the number of desired itemsets, without necessitating a minimum utility threshold. The existing top-$k$ HUIM algorithms can be classified into two main categories: two-stage algorithms (such as TKU \cite{wu2012TKU} and REPT \cite{ryang2015REPT}) and one-stage algorithms (such as TKO \cite{tseng2015TKOandTKU}, KHMC \cite{duong2016KHMC}, and THUI \cite{krishnamoorthy2019mining}). The top-$k$ HUIM holds significant importance and broad applicability across various domains, such as e-commerce \cite{song2021topums} and healthcare \cite{dong2019mining}. Although utility-based mining algorithms are widely adopted, considering only the utility of itemsets is insufficient, as some HUIs in the short term may not hold significant value in the long run. For example, certain products might generate high profits initially due to advertising and user curiosity, but their value may diminish over time as interest wanes. Therefore, periodic frequent pattern mining \cite{tanbeer2009periodic-frequent} was introduced to discover itemsets frequently observed over extended periods. However, merely identifying periodic frequent patterns is inadequate because this does not consider the importance of the itemsets. To overcome this challenge, periodic high-utility itemset mining (PHUIM) \cite{fournier2016phm} was introduced to identify itemsets that maintain high utility over the long term. Periodic pattern mining has widespread applications, such as traffic congestion analysis \cite{likitha2021discovering}, disaster event detection \cite{kiran2017discovering}, and marketing analysis \cite{lai2023PHMN}.

The tasks of PHUIM and top-$k$ HUIM have been widely applied in real-world scenarios. PHUIM requires users to predefine the \textit{minutil}, and the itemsets discovered by top-$k$ HUIM algorithms may not have long-term high utility. Setting an appropriate \textit{minutil} is a challenging task. Moreover, users are generally more interested in the top-$k$ periodic itemsets with the highest utility values than all itemsets that meet the \textit{minutil} and periodic constraints. Therefore, incorporating periodic constraints into the top-$k$ HUIM would yield more meaningful patterns. The top-$k$ periodic high-utility itemset mining (TPUM) has extensive practical applications, as it can uncover high-value patterns that rank in the top-$k$ in terms of utility and exhibit periodic characteristics. For instance, in the retail industry, businesses can optimize inventory management and promotional strategies by identifying customers' top-$k$ periodic purchasing behaviors, such as monthly purchases of high-profit product combinations. In the healthcare sector, analyzing patients' top-$k$ periodic visit patterns, such as regular follow-ups for high-frequency diseases or medication usage, can aid in disease prediction and treatment optimization. In finance, recognizing clients' top-$k$ periodic transaction behaviors, such as recurring investments in high-yield products or repayment patterns, can enhance risk management and personalized services. In intelligent transportation systems, mining top-$k$ periodic driving routes, such as high-traffic travel routes during peak hours, can support the prediction of traffic flow and route planning. In these scenarios, TPUM not only reveals the most valuable information but also ensures the security of sensitive data through privacy-preserving techniques such as data anonymization or differential privacy, thus achieving a balance between data utility and privacy protection.

Despite the widespread application of TPUM in practical applications, no research has specifically addressed this problem. Designing an efficient algorithm to mine this type of itemset presents significant challenges. The following discussion outlines the main challenges in solving this problem. First, a key challenge lies in effectively raising the threshold without overlooking potential itemsets. The threshold-increasing strategies employed in top-$k$ HUIM are not directly applicable to TPUM. This is because the strategies used in top-$k$ HUIM do not consider periodic factors when raising the \textit{minutil} threshold. Consequently, the threshold might be raised to an excessively high value, making it difficult to discover a sufficient number of PHUIs. Second, efficiently reducing the search space presents a critical challenge. The pruning strategies used in PHUIM often cannot be effectively employed in the initial stages. This is because, at the start, the \textit{minutil} is usually set to 0, allowing many itemsets to pass this condition. To summarize, this paper introduces an innovative TPUM framework designed to resolve the challenges mentioned earlier, where $k$ specifies the number of PHUIs to be identified. The primary contributions of this work are summarized as follows:

\begin{itemize}
    \item This work introduces a novel TPU algorithm specifically for mining top-$k$ PHUIs featuring both positive and negative utility values.
    
    \item A novel data structure, namely PEUCS, is designed to reduce the join operations of the utility list.
    
    \item We developed three novel threshold-raising strategies, PIU, PCUD, and PRU, to quickly elevate the threshold.
    
    \item The performance of the TPU algorithm is significantly improved by these threshold-raising strategies, as evidenced by the experimental results.
\end{itemize}

The remainder of this paper is organized as follows: Section \ref{sec: relatedwork} reviews existing research on HUIM, PHUIM, and top-$k$ HUIM. Section \ref{sec: preliminaries} introduces some key definitions related to top-$k$ PHUIs and the addressed data mining task. Section \ref{sec: algorithm} presents the proposed TPU algorithm. Section \ref{sec: experiments} discusses the results of the experimental evaluations. Finally, Section \ref{sec: conclusion} concludes with a summary and suggestions for future research.

\section{Related Work} \label{sec: relatedwork}
This section summarizes the literature on HUIM, PHUIM, and top-$k$ HUIM. Table \ref{table:HUIM} provides a detailed summary of various classic HUIM, PHUIM, and top-k HUIM algorithms.

\subsection{High-utility itemset mining}

\begin{table*}[h]
    \centering
    \caption{Comparison of HUIM, PHUIM, and top-$k$ HUIM algorithms and their innovations}
    \resizebox{\textwidth}{!}{
    \begin{tabular}{|c|c|c|c|c|c|c|}
    \hline
    \textbf{Classification} & \textbf{Algorithm} & \textbf{Year} & \textbf{Innovation} & \textbf{Periodicity} & \textbf{Top-$k$} & \textbf{Negative utility} \\ \hline
    \multirow{1}{*}{} & Two-Phase \cite{liu2005two-PhaseAlgorithm} & 2005 & TWDCP & $\times$ & $\times$ & $\times$ \\ \cline{2-7}
    &UP-Growth \cite{tseng2010UP-Growth} & 2010 & DGU, DGN, DLU, DLN & $\times$ & $\times$ & $\times$ \\ \cline{2-7}
    &UP-Growth+ \cite{tseng2012UP-GrowthAndUP-Growth+} & 2012 & DNU, DNN & $\times$ & $\times$ & $\times$ \\ \cline{2-7}
    HUIM&HUI-Miner \cite{liu2012mining} & 2012 & UL, RU & $\times$ & $\times$ & $\times$ \\ \cline{2-7}
    &FHM \cite{fournier2014fhm} & 2014 & EUCP & $\times$ & $\times$ & $\times$ \\ \cline{2-7}
    &FHN \cite{lin2016fhn} & 2016 & PNUL & $\times$ & $\times$ & $\checkmark$ \\ \cline{2-7}
    &R-Miner \cite{sra2024residual} & 2023 & residue-map, master-map & $\times$ & $\times$ & $\times$ \\ \cline{2-7}
    &FOTH \cite{yan2024efficient} & 2024 & propagation, IndexSet, mining area & $\times$ & $\times$ & $\times$ \\ \hline
    \multirow{1}{*}{PHUIM}&PHM \cite{fournier2016phm} & 2016 & \makecell*[c]{periodicity metrics, maxPer pruning, \\ maxAvg pruning, ESCS} & $\checkmark$ & $\times$ & $\times$ \\ \cline{2-7}
    &PHMN \cite{lai2023PHMN} & 2023 & M-list, DU strategy & $\checkmark$ & $\times$ & $\checkmark$ \\ \hline
    \multirow{1}{*}{}&TKU \cite{tseng2015TKOandTKU} & 2015 & PE, NU, MD, MC and SE & $\times$ & $\checkmark$ & $\times$ \\ \cline{2-7}
    &TKO \cite{tseng2015TKOandTKU} & 2015 & RUC, RUZ and EPB & $\times$ & $\checkmark$ & $\times$ \\ \cline{2-7}
    top-$k$ HUIM&REPT \cite{ryang2015REPT} & 2014 & RIU, PUD, RSD and SEP & $\times$ & $\checkmark$ & $\times$ \\ \cline{2-7}
    &kHMC \cite{duong2016KHMC} & 2016 & EUCPT, TEP, COV, CUD & $\times$ & $\checkmark$ & $\times$ \\ \cline{2-7}
    &THUI \cite{krishnamoorthy2019mining} & 2018 & LIU, LIU-E, LIU-LB & $\times$ & $\checkmark$ & $\times$ \\ \hline
    \multirow{1}{*}{top-$k$ PHUIM} & TPU (ours) & 2025 & PIU, PCUD, PRU & $\checkmark$ & $\checkmark$ & $\checkmark$ \\ \hline
    \end{tabular}
    }
    \label{table:HUIM}
\end{table*}

Pattern mining \cite{agrawal1994Apriori,han2000FP-Growth} is widely used in real-life applications, but considering only itemset support is insufficient. To overcome this issue, Chan et al. \cite{chan2003mining} introduced a new research problem called HUIM. With advancements in science and technology, many HUIM algorithms have been proposed \cite{gan2020survey,cheng2023efficient,li2023fchm,qu2023mining}. These algorithms fall into three main categories: two-phase algorithms, tree-based algorithms, and utility-list-based algorithms. \textbf{Two-phase algorithms:} To discover HUIs efficiently, Liu et al. \cite{liu2005two-PhaseAlgorithm} proposed the Two-Phase algorithm.  This algorithm is notable for introducing the transaction weight utility (TWU) model and the downward closure property (TWDCP). This innovation primarily consists of two stages for mining HUIs. However, these algorithms have two critical drawbacks. First, TWDCP is a loose upper bound, which generates many candidates. Second, repeated database scans are required. \textbf{Tree-based algorithms:} Due to the relatively loose TWDCP strategy used in two-phase algorithms, many candidates are generated. To tackle this problem, tree-based algorithms have been introduced. Tseng et al. \cite{tseng2012UP-GrowthAndUP-Growth+,tseng2010UP-Growth} developed two algorithms, UP-Growth and UP-Growth+, for mining HUIs. They utilized the UP-Tree \cite{han2000FP-Growth} and designed several strategies, such as DGU, DGN, DLU, DLN, DNU, and DNN, to effectively reduce the overestimated utility and the count of candidate itemsets. The tree-based algorithm framework consists of three steps: first, constructing a UP-Tree; then, using the corresponding algorithm to generate potential high-utility itemsets (pt-HUIs) from the UP-Tree; and ultimately, determining the true HUIs from the generated set of pt-HUIs. Zhang et al. \cite{zhang2023mining} introduced the FHUSN algorithm to mine high-utility sequences under scenarios that may or may not include negative utility value. \textbf{Utility-list-based algorithms:} Liu et al. \cite{liu2012mining} developed an algorithm known as HUI-Miner for efficiently mining HUIs. Its innovation lies in introducing a novel utility list (UL) and a new remaining utility (RU) pruning strategy. The UL data structure allows mining HUIs without generating candidate itemsets. However, HUI-Miner requires costly UL join operations. The novel FHM algorithm \cite{fournier2014fhm} integrates a novel strategy called EUCP, effectively reducing the number of UL join operations during HUIM. Lin et al. \cite{lin2016fhn} introduced an algorithm named FHN, aiming to find HUIs with negative items efficiently. Its innovation lies in designing a vertical list structure called the PNUL. Compared to tree-based algorithms, utility-list-based one-phase algorithms demonstrate superior performance. Several cutting-edge methods for mining HUIs have been introduced. The FOTH algorithm \cite{yan2024efficient} employs a unique structure called IndexSet to efficiently mine HUIs, bypassing join operations by propagating IndexSets. The R-Miner algorithm \cite{sra2024residual} relies on two sophisticated data structures (the residue map and the master map) to accurately discover HUIs. For handling large-scale datasets, a GPU-accelerated heuristic algorithm, PHA-HUIM \cite{fang2023gpu}, has been proposed. Additionally, a novel parallelized approach, DPHIM \cite{2023HUIOnMulticoreProcessors}, has been developed, providing an effective and scalable solution for HUI mining. 

\subsection{Periodic high-utility itemset mining}

Although the algorithms mentioned above can be effectively applied to HUIM, considering only the utility of itemsets during HUI mining is insufficient because some itemsets with high utility in the short term may not possess significant value in the long term. Zhang et al. \cite{zhang2022fuzzy} introduced two algorithms, FP2M and SFP2M, for discovering periodic frequent patterns. Fournier-Viger et al. \cite{fournier2016phm} introduced PHUIM and developed an algorithm named PHM to identify itemsets that maintain high utility over extended periods. PHUIM not only considers the utility of itemsets in transaction databases but also takes into account the periodicity of itemsets. The PHM algorithm utilizes a utility list structure \cite{fournier2014fhm,liu2012mining} and introduces new metrics for pattern periodicity, namely average periodicity (\textit{avgPer}) and minimum periodicity (\textit{minPer}), to flexibly assess the periodicity of patterns. Additionally, two new pruning strategies, namely maximum periodicity (\textit{maxPer}) pruning and maximum average periodicity (\textit{maxAvg}) pruning, are proposed. Furthermore, two new optimization strategies, namely ALC pruning and EAPP pruning based on the proposed ESCS structure, are introduced. Qi et al. \cite{qi2023mining} introduced the CPR-Miner algorithm, which is the first to focus on mining closed periodic recent HUPs. Lai et al. \cite{lai2023PHMN} proposed the PHMN algorithm to discover PHUIs in databases with negative items. Their innovation includes the introduction of a novel data structure, called the M-list, and the development of an effective upper bound along with a DU strategy tailored to this structure, which enhances the efficiency of search space pruning. Following this, they introduced PHMN+, an enhanced version of the original algorithm that utilizes the DU pruning technique. 

\subsection{Top-$k$ high-utility itemset mining}

Mining HUIs requires users to predefine a \textit{minutil} value, but setting an appropriate \textit{minutil} value is challenging. Consequently, studies proposed the top-$k$ HUIM problem, which doesn't require a predefined \textit{minutil} value. Instead, it mines HUIs based on a user-specified value of $k$, representing the number of desired HUIs. Existing top-$k$ HUIM algorithms can be broadly divided into two categories: two-phase and one-phase methods. In the two-phase approach, potential top-$k$ HUIs are first generated, and then the top-$k$ HUIs are identified from these results in the subsequent phase. On the other hand, one-phase algorithms are designed to find the top-$k$ HUIs directly without the need for multiple database scans. Tseng et al. \cite{tseng2015TKOandTKU} introduced two efficient algorithms, TKU and TKO, which are specifically tailored to mine the full set of top-$k$ HUIs. The TKU algorithm is a two-phase approach that uses strategies such as PE, NU, MD, MC, and SE to raise \textit{minutil} and prune the search space. The TKO algorithm is a one-phase method that employs strategies like RUC, RUZ, and EPB to improve performance. The UP-Tree structure \cite{tseng2010UP-Growth} is utilized by the TKU algorithm for maintaining itemset utility and transaction details, while the UL structure \cite{liu2012mining} is employed by the TKO algorithm to keep itemset utility information. Ryang et al. \cite{ryang2015REPT} introduced the REPT algorithm, which effectively reduces the number of generated candidate itemsets. REPT uses the UP-Tree  \cite{tseng2012UP-GrowthAndUP-Growth+,tseng2010UP-Growth} and employs strategies such as RIU, PUD, RSD, and SEP to raise the threshold. The kHMC algorithm \cite{duong2016KHMC} utilizes the EUCPT pruning strategy to reduce many join operations and introduces the TEP strategy to prune itemsets and their extensions before constructing UL. Additionally, the kHMC algorithm utilizes RIU \cite{ryang2015REPT}, COV, and CUD threshold-raising techniques to approach the optimal threshold value. In the THUI algorithm \cite{krishnamoorthy2019mining}, its innovation lies in introducing the LIU data structure, along with the novel strategies LIU-E and LIU-LB. Gan et al. \cite{gan2020tophui} introduced the TopHUI algorithm, which utilizes the PNUL data structure \cite{lin2016fhn} to reduce database accesses. The TOPIC algorithm \cite{chen2021topic} was proposed to overcome the time and memory inefficiencies associated with the TopHUI algorithm \cite{gan2020tophui}. Song et al. \cite{song2021topums} introduced TopUMS, an algorithm designed to extract top-$k$ HUIs from data streams using a sliding window approach. Zhang et al. \cite{zhang2021tkus} defined the top-$k$ HUSPM problem and also developed the TKUS algorithm. Chen et al. \cite{chen2022generic} developed the TOIT method for discovering top-$k$ on-shelf HUIs. The TMKU algorithm \cite{huang2023targeted} tackles the challenge of finding targeted top-$k$ HUIs, while the threshold-raising strategies employed in top-$k$ HUIM are not directly applicable to TPUM. This is because these strategies raise the \textit{minutil} threshold without considering periodicity, which may result in the threshold being set too high. Consequently, this can lead to the failure of mining the desired top-$k$ PHUIs.

\section{Preliminaries} \label{sec: preliminaries}

In this section, we define the essential concepts of HUIM and formally introduce the TPUM problem. Let $I$ = \{$i_{1}$, $i_{2}$, $i_{3}$, $\cdots$, $i_{m}$\} denote the collection of items and $\mathcal{D}$ represents a database. Here, $m$ represents the total count of items in $\mathcal{D}$. In general, $\mathcal{D}$ consists of a utility table and a set of transactions. The utility table records the utility of each item in $I$. Each transaction $T$ in the transaction database is assigned a unique identifier, \textit{tid}. Each transaction comprises a collection of items, $X$, where $X$ $\subseteq$ $I$. Each itemset $Z$ is then a nonempty subset of the transaction $X$, where $Z$ $\subseteq$ $X$. Each item $x$ $\in$ $T_i$ is associated with an internal utility value, representing its quantity. Additionally, each item $x$ in $T_i$ is assigned an external utility value, representing its profit, price, or others, as shown in Table \ref{tab:external_utility}. If an itemset $Z$ contains $k$ items, it is termed as an $k$-itemset. For instance, in Table \ref{Database}, this transaction database comprises ten transactions, which are composed of several items: $a$, $b$, $c$, $d$, $e$, $f$, $g$. Table \ref{tab:external_utility} displays the external utility associated with each item. For instance, the transaction $T_5$ comprises five items: $c$, $d$, $e$, $f$, and $g$, with internal utilities of 5, 1, 1, 2, and 3, respectively. Besides, the external utilities for these items $c$, $d$, $e$, $f$, and $g$ are -1, 3, -2, 6, and -1, respectively.

\begin{table}[h]
    \centering
    \caption{An example database}
    \begin{tabular}{|l|c|l|}
    \hline
    \textbf{Tid} & \textbf{Transaction \{item:quantity\}} & \textbf{TU} \\
    \hline
    $T_{1}$ & \{$a$:1\} \{$c$:1\} \{$d$:3\} \{$e$:2\} \{$g$:1\} & 5 \\
    \hline
    $T_{2}$ & \{$a$:3\} \{$b$:1\} \{$e$:5\} \{$f$:7\}  &  43  \\
    \hline
    $T_{3}$ &  \{$b$:1\} \{$c$:3\} \{$d$:4\} \{$f$:2\} \{$g$:6\}  & 20    \\
    \hline
    $T_{4}$ & \{$a$:1\} \{$d$:2\} \{$e$:3\} & 2 \\
    \hline
    $T_{5}$ & \{$c$:5\} \{$d$:1\} \{$e$:1\} \{$f$:2\} \{$g$:3\} &  5   \\
    \hline
    $T_{6}$ &  \{$b$:2\} \{$c$:1\} \{$f$:5\}  & 39    \\
    \hline
    $T_{7}$ & \{$a$:3\} \{$d$:2\} \{$g$:2\} & 10 \\
    \hline
    $T_{8}$ & \{$a$:1\} \{$b$:2\} \{$c$:1\} \{$d$:1\} \{$e$:2\} \{$f$:1\} \{$g$:1\} & 15  \\
    \hline
    $T_{9}$ &  \{$c$:3\} \{$d$:4\} \{$f$:5\} & 39    \\
    \hline
    $T_{10}$ & \{$a$:1\} \{$b$:1\} \{$c$:2\} \{$d$:1\} \{$e$:2\}  & 4  \\
    \hline
    \end{tabular}
    \label{Database}
\end{table}

\begin{table}[h]
    \centering
    \caption{Utility table}
    \begin{tabular}{|l|l|l|l|l|l|l|l|}
        \hline
        \textbf{Item} & \textbf{$a$} & \textbf{$b$} & \textbf{$c$} & \textbf{$d$} & \textbf{$e$} & \textbf{$f$} & \textbf{$g$} \\
        \hline
        \textbf{Utility} & 2 & 5 & -1 & 3 & -2 & 6 & -1 \\
        \hline
        \end{tabular}
    \label{tab:external_utility}
\end{table}

\begin{definition}[utility of the itemset]
    \rm In $\mathcal{D}$, each item $x$ $\in$ $I$ is assigned an external utility \textit{eu}$(x)$. In $T_i$, each item $x$ possesses an internal utility \textit{iu}$(x,T_i)$. The utility of $x$ in $T_i$ is $U(x, T_i)$ = \textit{iu}$(x,T_i)$ $\times$ \textit{eu}$(x)$. For an itemset $X$ and transaction $T$, the positive utility \textit{PU}$(X,T)$ and negative utility \textit{NU}$(X,T)$ are \textit{PU}$(X,T)$ = $\sum_{i \in X \wedge eu(i) > 0}$ $U(i,T)$ and \textit{NU}$(X,T)$ = $\sum_{i \in X \wedge eu(i) < 0}$ $U(i,T)$. In $\mathcal{D}$, the positive utility \textit{PU}$(X)$ and negative utility \textit{NU}$(X)$ of itemset $X$ are \textit{PU}$(X)$ = $\sum_{X \subseteq T_j \subseteq \mathcal{D}}$ \textit{PU}$(X,T_j)$ and \textit{NU}$(X)$ = $\sum_{X \subseteq T_j \subseteq \mathcal{D}}$ \textit{NU}$(X,T_j)$. The utility of itemset $X$ is $U(X)$ = \textit{PU}$(X)$ + \textit{NU}$(X)$.
\end{definition}

For instance, in Table \ref{Database} and Table \ref{tab:external_utility}, \textit{iu}$(e,T_5)$ = 1, \textit{eu}$(e)$ = -2, and $U(e,T_5)$ = \textit{iu}$(e,T_5)$ $\times$ \textit{eu}$(e)$ = 1 $\times$ (-2)  = -2. Let itemset $X$ = $bc$. We have \textit{PU}$(X)$ = \textit{PU}$(X,T_3)$ + \textit{PU}$(X,T_6)$ + \textit{PU}$(X,T_8)$ + \textit{PU}$(X,T_{10})$ = \textit{PU}$(b,T_3)$ + \textit{PU}$(b,T_6)$ + \textit{PU}$(b,T_8)$ + \textit{PU}$(b,T_{10})$ = 5 + 10 + 10 + 5 = 30, \textit{NU}$(X)$ = \textit{NU}$(X,T_3)$ + \textit{NU}$(X,T_6)$ + \textit{NU}$(X,T_8)$ + \textit{NU}$(X,T_{10})$ = \textit{NU}$(c,T_3)$ + \textit{NU}$(c,T_6)$ + \textit{NU}$(c,T_8)$ + \textit{NU}$(c,T_{10})$ = -7, and $U(X)$ = $U(X,T_3)$ + $U(X,T_6)$ + $U(X,T_8)$ + $U(X,T_{10})$ = 2 + 9 + 9 + 3 = 23. Notice that the equation $U(X)$ = \textit{PU}$(X)$ + \textit{NU}$(X)$ holds for any itemset $X$.

\begin{definition}[redefined transaction-weighted utility \cite{lin2016fhn}]
     \rm Within $\mathcal{D}$, the transaction utility \textit{TU}$(T_k)$ equals the total utility of all items in $T_k$ and \textit{TU}$(T_k)$ = $\sum_{i \in T_k}$ $U(i,T_k)$. The redefined transaction utility of $T_k$, \textit{RTU}$(T_k)$, equals the total utility of all items in $T_k$ with positive external utility and \textit{RTU}$(T_k)$ = \textit{PU}$(T_k, T_k)$. For an itemset $X$, the transaction-weighted utility in $\mathcal{D}$, \textit{TWU}$(X)$, is the sum of the TU of all transactions containing $X$ and \textit{TWU}$(X)$ = $\sum_{X \subseteq T_k \subseteq \mathcal{D}}$ \textit{TU}$(T_k)$. For an itemset $X$, the redefined transaction-weighted utility in $\mathcal{D}$, \textit{RTWU}$(X)$, is the sum of the RTU of all transactions containing $X$ and \textit{RTWU}$(X)$ = $\sum_{X \subseteq T_k \subseteq \mathcal{D}}$ \textit{RTU}$(T_k)$.
\end{definition}

For instance, the \textit{TU} and \textit{RTU} for transaction $T_i$ (1 $\leqslant$ $i$ $\leqslant$ 10) are illustrated in Table \ref{tab:TUandRTU}.  Table \ref{tab:TWUandRTWU} displays the values of \textit{TWU} and \textit{RTWU} for all single items. The values of \textit{TU} and \textit{RTU} for the same transaction are different, as are the values of \textit{TWU} and \textit{RTWU} for the same item. This can be easily observed from Table \ref{tab:TUandRTU} and Table \ref{tab:TWUandRTWU}: \textit{TU}$(T_{1})$ = 5, \textit{RTU}$(T_{1})$ = 11, \textit{TWU}$(a)$ = 79, and \textit{RTWU}$(a)$ = 115.

\begin{table}[h]
    \centering
    \small
    \caption{TU and RTU of each transaction}
    \begin{tabular}{|c|c|c|c|c|c|c|c|c|c|c|}
        \hline
        \textbf{\textit{Tid}} & $T_{1}$ & $T_{2}$ & $T_{3}$ & $T_{4}$ & $T_{5}$ & $T_{6}$ & $T_{7}$ & $T_{8}$ & $T_{9}$ & $T_{10}$ \\
        \hline
        \textbf{TU} & 5 & 43 & 20 & 2 & 5 & 39 & 10 & 15 & 39 & 4 \\
        \hline
        \textbf{RTU} & 11 & 53 & 29 & 8 & 15 & 40 & 12 & 21 & 42 & 10 \\
        \hline
        \end{tabular}
    \label{tab:TUandRTU}
\end{table}

\begin{table}[h]
    \centering
    \caption{TWU and RTWU of each item}
    \begin{tabular}{|c|c|c|c|c|c|c|c|}

        \hline
        \textbf{Item} & $a$ & $b$ & $c$ & $d$ & $e$ & $f$ & $g$ \\ 
        \hline
        \textbf{TWU} & 79 & 121 & 127 & 100 & 74 & 161 & 55 \\
        \hline
        \textbf{RTWU} & 115 & 153 & 168 & 148 & 118 & 200 & 88 \\
        \hline
        \end{tabular}
    \label{tab:TWUandRTWU}
\end{table}

\begin{definition}[top-$k$ high utility itemset \cite{tseng2015TKOandTKU}]
    \rm Let \textit{minutil} denotes the user-defined minimum utility threshold. If $U(X)$ $\geq$ \textit{minutil} for an itemset $X$, then $X$ is termed as a HUI. In contrast, if $U(X)$ $<$ \textit{minutil}, then $X$ is a low-utility itemset. If there are at least $k$ itemsets with utility no less than the \textit{minutil}, then the top-$k$ HUIs are the $k$ itemsets with the highest utility.
\end{definition}

For example, if $k$ = 3 and \textit{minutil} = 90, then from Table \ref{Database}, we can observe that the top-3 HUIs are $\{(d,f):90\}$, $\{(b,f):120\}$, and $\{(f):132\}$.

\begin{definition}[itemset period \cite{fournier2016phm,lai2023PHMN,tanbeer2009periodic-frequent}]
    \label{definition:ItemsetPeriod}
    \rm Given an itemset $X$, all transactions containing $X$ can be obtained by scanning $\mathcal{D}$ once. The set of transactions can be represented by their transaction \textit{tid} as $T(X)$ = $\{tid | X \subseteq T_{tid}\}$ = \{$tid_{1}$, $tid_{2}$, $\cdots$, $tid_{n}$\}, where $n$ $\leqslant$ $|\mathcal{D}|$. Here, $|\mathcal{D}|$ denotes the size of the database $\mathcal{D}$. The period of two consecutive transactions is the difference between their identifiers $tid$, i.e., the $tid$ of the later appearing transaction minus the $tid$ of the first appearing transaction. The period of an itemset $X$ is the adjacency interval of the consecutive transactions that contain $X$. We can define the periods of an itemset $X$, \textit{Per}$(X)$ = \{$tid_{i+1}$ - $tid_{i}$ $|$ $i$ = 0, 1, 2, $\cdots$ $k$\}, where $tid_{0}$ = 0 and $tid_{i+1}$ = $|\mathcal{D}|$. Using \textit{Per}$(X)$, we can precisely determine the maximum time interval, minimum time interval, and average time interval for the appearance of $X$, denoted as \textit{maxPer}$(X)$, \textit{minPer}$(X)$, and \textit{avgPer}$(X)$, respectively.  \textit{maxPer}$(X)$ = \textit{max(Per(X))}, \textit{minPer}$(X)$ = \textit{min(Per(X))}, \textit{avgPer}$(X)$ = $\frac{( \sum_{k \in Per(X)} k)}{|Per(X)|} $ = $\frac{|\mathcal{D}|}{ (|sup(X)| + 1)} $. Here, $|$\textit{sup(X)}$|$ represents the support of $X$, i.e., the total transactions that contain $X$. It is important to note that when calculating \textit{minPer}$(X)$, we exclude the first and last elements of \textit{Per}$(X)$. Detailed explanations for this exclusion can be found in Ref. \cite{fournier2016phm}.
\end{definition}

For instance, for the item $e$ in Table \ref{Database}, we can calculate that $T(e)$ = \{1, 2, 4, 5, 8, 10\} and \textit{Per}$(e)$ = \{1, 1, 2, 1, 3, 2, 0\}. Consequently, we can easily derive that \textit{maxPer}$(e)$ = \textit{max(Per(e))} = 3, \textit{minPer(e)} = \textit{min(Per(e))} = 1, \textit{avgPer(e)} = $\frac{ |\mathcal{D}| }{ (|sup(X)| + 1) }$ = $\frac{10}{( 6+1 )}$ = 1.429. The periodic information for all single items is summarized in Table \ref{table:singleitemperiodic}.

\begin{table}[h]
    \centering
    \caption{The periodic and support information of each item}
    \begin{tabular}{|c|c|c|c|c|c|c|c|}
    \hline
    \textbf{Item} & $a$ & $b$ & $c$ & $d$ & $e$ & $f$ & $g$ \\ \hline
    \textbf{\textit{maxPer}}  & 3 & 3 & 2 & 2 & 3 & 2 & 2 \\ \hline
    \textbf{\textit{minPer}} & 1 & 1 & 1 & 1 & 1 & 1 & 1 \\ \hline
    \textbf{\textit{support}} & 6 & 5 & 7 & 8 & 6 & 6 & 5 \\ \hline
    \textbf{\textit{avgPer}}  & 1.43 & 1.67 & 1.25 & 1.11 & 1.43 & 1.43 & 1.67 \\ \hline
    \end{tabular}
    \label{table:singleitemperiodic}
\end{table}

\begin{definition}[periodic high utility itemset \cite{fournier2016phm}]
    \rm Before analysis, users define the periodicity thresholds and \textit{minutil}, denoted as \textit{maxPer}, \textit{minPer}, \textit{maxAvg}, \textit{minAvg}, and \textit{minutil}, for the itemsets of interest. For any itemset $X$, if the conditions  \textit{minPer} $\leqslant$ \textit{minPer}$(X)$, \textit{maxPer}$( X)$ $\leqslant $ \textit{maxPer}, \textit{minAvg} $\leqslant$ \textit{avgPer}$(X)$ $\leqslant$ \textit{maxAvg}, and $U(X)$ $\geqslant$ \textit{minutil} hold true, $X$ is thus classified as a PHUI.
\end{definition}

For example, if \textit{minutil} = 70, \textit{minPer} = 1, \textit{maxPer} = 5, \textit{minAvg} = 1, and \textit{maxAvg} = 5, then from Table \ref{Database}, all PHUIs that meet these criteria can be found as shown in Table \ref{table:PHUIs}.

\begin{table}[h]
    \centering
    \small
    \caption{The set of PHUIs in the running example}
    \begin{tabular}{|c|c|c|c|c|}
    \hline
    \textbf{Itemset} & \textbf{\textit{U(X)}} & \textbf{\textit{minPer}}$(X)$ & \textbf{\textit{maxPer}}$(X)$ & \textbf{\textit{avgPer}}$(X)$ \\ \hline
    $\{c, d, f\}$     & 78  & 1  & 3  & 2.0 \\ \hline
    $\{c, f\}$  & 77  & 1  & 3  & 1.7 \\ \hline
    $\{d, f\}$ & 90 & 1  & 3  & 2.0 \\ \hline
    $\{b, f\}$  & 120  & 1  & 3  & 2.0 \\ \hline
    $\{f\}$     & 132  & 1  & 2  & 1.4  \\ \hline
    \end{tabular}
    \label{table:PHUIs}
\end{table}

\begin{definition}[order of items \cite{lai2023PHMN}]
    \label{OrderofItems}
    \rm Given a set of items $I^{\star}$ = \{$i_{1}$, $i_{2}$, $i_{3}$, $\cdots$, $i_{k}$\}, an order $(\prec)$ can be defined on these items. For any two distinct items $i_{p}$, $i_{q}$ $\in$ $I^{\star}$, if both $i_{p}$ and $i_{q}$ are either positive or negative items, we will arrange them in ascending order according to their \textit{RTWU} values. If $i_{p}$ is a positive item and $i_{q}$ is a negative item, then $i_{q}$ $\prec$ $i_{p}$.
\end{definition}

From Table \ref{tab:TWUandRTWU}, we can observe the \textit{RTWU} values for all items. From Table \ref{tab:external_utility}, we can identify $c$, $e$, and $g$ as the negative items. Therefore, the order of the items is \{$g$ $\prec$ $e$ $\prec$ $c$ $\prec$ $a$ $\prec$ $d$ $\prec$ $b$ $\prec$ $f$\}.

\textbf{Problem statement.} Given a transactional database $\mathcal{D}$, periodicity constraints \textit{minPer}, \textit{maxPer}, \textit{minAvg}, \textit{maxAvg}, and the number $k$ of PHUIs, the problem of TPUM is to discover the top-$k$ itemsets with the highest utility that also satisfy the periodic constraints from $\mathcal{D}$. The TPUM task is illustrated through an example. Given $\mathcal{D}$ in Table \ref{Database}, with \textit{minPer} = 1, \textit{maxPer} = 5, \textit{minAvg} = 1, \textit{maxAvg} = 5 and $k$ = 3, we can obtain the top-3 PHUIs from $\mathcal{D}$ as $\{(d,f)$: 90\}, $\{(b,f)$: 120\}, and $\{(f)$: 132\}. The discovered top-3 periodic high-utility patterns are shown in Table \ref{table:PHUIs}.

\section{Proposed Algorithm} \label{sec: algorithm}

In this section, we introduce the data structures, pruning strategies, and threshold-raising strategies used in this paper. Furthermore, we propose a novel TPU algorithm for discovering top-$k$ PHUIs in databases that may or may not include negative utility values.

\begin{table}[h]
    \centering
    \caption{A reordered database}
    \label{Reordered_Database}
    \begin{tabular}{|c|c|c|}
        \hline
        \textbf{Tid} & \textbf{Transaction\{item:utility\}} & \textbf{RTU} \\
        \hline
        $T_{1}$ & \{$g$:-1\} \{$e$:-4\} \{$c$:-1\} \{$a$:2\} \{$d$:9\} & \$11 \\
        \hline
        $T_{2}$ & \{$e$:-10\} \{$a$:6\} \{$b$:5\} \{$f$:42\} & \$53 \\
        \hline
        $T_{3}$ & \{$g$:-6\} \{$c$:-3\} \{$d$:12\} \{$b$:5\} \{$f$:12\} & \$29 \\
        \hline
        $T_{4}$ & \{$e$:-6\} \{$a$:2\} \{$d$:6\} & \$8 \\
        \hline
        $T_{5}$ & \{$g$:-3\} \{$e$:-2\} \{$c$:-5\} \{$d$:3\} \{$f$:12\} & \$15 \\
        \hline
        $T_{6}$ & \{$c$:-1\} \{$b$:10\} \{$f$:30\} & \$40 \\
        \hline
        $T_{7}$ & \{$g$:-2\} \{$a$:6\} \{$d$:6\} & \$12 \\
        \hline
        $T_{8}$ & \{$g$:-1\} \{$e$:-4\} \{$c$:-1\} \{$a$:2\} \{$d$:3\} \{$b$:10\} \{$f$:6\} & \$21 \\
        \hline
        $T_{9}$ & \{$c$:-3\} \{$d$:12\} \{$f$:30\} & \$42 \\
        \hline
        $T_{10}$ & \{$e$:-4\} \{$c$:-2\} \{$a$:2\} \{$d$:3\} \{$b$:5\} & \$10 \\
        \hline 
    \end{tabular}
\end{table}

\subsection{Data structures}

\begin{definition}[extension]
    \rm For each item in an ordered itemset, suppose the current itemset is \{$p_1$, $p_2$, $\cdots$, $p_n$\}, then each item $p_i$ can be extended to combinations that include $p_i$ and all subsequent items that have not yet appeared.
\end{definition}

For example, consider the ordered itemset \{$a$, $b$, $c$, $d$\}. Then, $a$ can be extended to $ab$, $ac$, $ad$. The $ab$ can be extended to $abc$ and $abd$. The $abc$ can be extended to $abcd$. This process continues, but $d$ cannot be extended further.

\begin{definition}[PNU-list \cite{lin2016fhn}]
    \rm Given an itemset $X$ in $\mathcal{D}$, the PNU-list of $X$, denoted as \textit{X.PNUL}, consists of two parts: the itemset $X$ itself and a list associated with $X$. This list comprises all transactions containing $X$, represented as a 4-tuple \textit{[tid}, \textit{pu}, \textit{nu}, \textit{ru]}, where \textit{tid} is the transaction identifier, \textit{pu} is the positive utility of $X$ in $T_{tid}$, \textit{nu} is the negative utility of $X$ in \textit{tid}, and \textit{ru} is the sum of the utilities of all positive items following $X$ in \textit{tid}. The value of \textit{ru} can be calculated as  \textit{ru}$(X, T_{tid})$ = $\sum_{i \in T_{tid} \wedge  x \prec i \wedge  \forall x \in X}$ \textit{PU}$(i, T_{tid})$.
\end{definition}

\begin{table}[ht]
    \centering
    \caption{PNU-list of \{$c$, $d$, $b$\}}.
    \begin{tabular}{cccc}
        \toprule
        \multicolumn{4}{c}{\textbf{Itemset \{$c$, $d$, $b$\}}} \\
        \midrule
        \textit{tid} & \textit{pu} & \textit{nu} & \textit{ru} \\
        3 & 17 & -3 & 12 \\
        8 & 13 & -1 & 6 \\
        10 & 8 & -2 & 0 \\
        \bottomrule
    \end{tabular}
    \label{PNU-list}
\end{table}

For example, the PNU-list structure of \{$c$,$d$,$b$\} is shown in Table \ref{PNU-list}. The Estimated Utility Co-occurrence Structure (EUCS) \cite{fournier2014fhm} is a 3-tuple defined as [$a$, $b$, $c$] $\in$ $I$ $\times$ $I$ $\times$ $R$, where $a$ and $b$ represent two distinct items, and $c$ represents the value of \textit{RTWU}$(a, b)$. For instance, the EUCS of items sorted according to Definition \ref{OrderofItems} is shown in Table \ref{table: EUCS}.

\begin{table}[h]
    \centering
    \caption{The EUCS of the running example}
    \begin{tabular}{|c|c|c|c|c|c|c|c|}
    \hline
    Item & $e$ & $c$ & $a$ & $d$ & $b$ & $f$\\ \hline
    $g$  & 47  & 76  & 44 & 88 & 50 & 65 \\ \hline
    $e$  &  & 57  & 103 & 65 & 84 & 89 \\ \hline
    $c$  &  & & 42 & 128 & 100 & 147 \\ \hline
    $a$  &  & & & 62 & 84 & 74 \\ \hline
    $d$  &  & & &  & 60 & 107 \\ \hline
    $b$  &  & & & &  & 143 \\ \hline
    \end{tabular}
    \label{table: EUCS}
\end{table}

\begin{definition}
    \rm The period estimated utility co-occurrence structure (PEUCS) is a 3-tuple represented as [$x$, $y$, $z$]. Here, $x$ and $y$ are two different items. $z$ contains the \textit{RTWU}$(x, y)$, $U(xy)$, and the periodicity information of the itemset \{$x$, $y$\}.
\end{definition}

Here, we provide a brief description of the construction process for PEUCS. First, we scan the first transaction $T_{1}$ in Table \ref{Reordered_Database}, which contains five items: $g$, $e$, $c$, $a$, and $d$. Consequently, PEUCS stores 2-itemsets: $ge$, $gc$, $ga$, $gd$, $ec$, $ea$, $ed$, $ca$, $cd$, and $ad$, with the utility of -5, -2, 1, 8, -5, -2, 5, 1, 8, and 11, respectively. When calculating \textit{minPer}, we do not consider the first and last period; therefore, only the \textit{maxPer} and \textit{support} of these itemsets are updated. The \textit{support} for all these itemsets is 1, and \textit{maxPer} is also set to 1. Next, we process the second transaction $T_{2}$. The 2-itemsets in $T_{2}$ are $ea$, $eb$, $ef$, $ab$, $af$, and $bf$, with the utility of -4, -5, 32, 11, 48, and 47, respectively. Their support increases by 1. Except for the itemset $ea$, all other itemsets appear for the first time, so their \textit{minPer} is not updated as explained earlier, and their \textit{maxPer} is set to 2. Since $ea$ already exists in PEUCS with a utility of -2, its utility is updated to -2 - 4 = -6, and \textit{minPer}$(ea)$ is set to 1. The difference between the $tid$ of $T_{2}$ and $T_{1}$ is 1, which is not greater than the existing \textit{maxPer}$(ea)$ in PEUCS, so \textit{maxPer}$(ea)$ is not updated. The same method is applied to process other transactions. The final results of the PEUCS for two different items are shown in Table \ref{table:PEUCS}.  Each entry in Table \ref{table:PEUCS} records the data from left to right, representing the \textit{RTWU}, \textit{Utility}, \textit{minPer}, \textit{maxPer}, and \textit{support} of all 2-itemsets. Note that in PEUCS, $^*$ indicates utility less than 0, while $^{\varDelta}$ indicates the itemset that does not satisfy Property \ref{maxPer_Pruning}.

\begin{table*}[h]
    \centering
    \small
    \caption{The PEUCS of the running example}
    \begin{tabular}{|c|c|c|c|c|c|c|}
    \hline
    \textbf{Item} & $e$ & $c$ & $a$ & $d$ & $b$ & $f$\\ \hline
    $g$     & 47, -15$^*$, 3, 4, 3  & 76, -21$^*$, 2, 3, 4  & 44, 6, 1, 6$^{\varDelta}$, 3 & 88, 20, 1, 2, 5 & 50, 8, 5, 5, 2 & 65, 20, 2, 3, 3 \\ \hline
    $e$   &  & 57, -23$^*$, 2, 4, 4  & 103, -14$^*$, 1, 4, 5 & 65, 4, 1, 3, 5 & 84, 2, 2, 6$^{\varDelta}$, 3 & 89, 44, 3, 3, 3 \\ \hline
    $c$ &   & & 42, 2, 2, 7$^{\varDelta}$, 3 & 128, 27, 1, 3, 6 & 100, 23, 2, 3, 4 & 147, 77, 1, 3, 5 \\ \hline
    $a$    &  & & & 62, 41, 1, 3, 5 & 84, 30, 2, 6$^{\varDelta}$, 3 & 74, 56, 6, 6$^{\varDelta}$, 2 \\ \hline
    $d$     &  & & &  & 60, 38, 2, 5, 3 & 107, 90, 1, 3, 4 \\ \hline
    $b$  &  & & & &  & 143, 120, 1, 3, 4 \\ \hline
    \end{tabular}
    \label{table:PEUCS}
\end{table*}

\subsection{Pruning strategies}

\begin{property}[RTWU pruning \cite{liu2005two-PhaseAlgorithm}]
    \label{RTWUPruning}
    \rm Let $X$ be an itemset. If \textit{RTWU}$(X)$ $<$ \textit{minutil}, then both $X$ and all its super-itemsets are considered low utility itemsets. Conversely, if \textit{RTWU}$(X)$ $\geq$ \textit{minutil}, then itemset $X$ passes the \textit{RTWU} pruning strategy and can be extended.
\end{property}
\textbf{Proof:} Let $X$ and $Y$ be two itemsets such that $X$ $\subseteq$ $Y$. We can easily derive the formulas \textit{RTWU}$(X)$ $\geqslant$  \textit{RTWU}$(Y)$ $>$ $U(Y)$ and \textit{RTWU}$(X)$ $>$ \textit{U}$(X)$. If \textit{RTWU}$(X)$ $<$ \textit{minutil}, then $U(X)$ $<$ \textit{minutil} and $U(Y)$ $<$ \textit{minutil}. Therefore, we can conclude that neither $X$ nor $Y$ are HUIs.

\begin{property}[RU pruning \cite{lin2016fhn,liu2012mining}]
    \label{RU_Pruning}
    \rm Given an itemset $X$, if \textit{RU}$(X)$ + \textit{PU}$(X)$ $<$ \textit{minutil}, then neither $X$ nor its extensions are HUIs. Conversely, if \textit{RU}$(X)$ + \textit{PU}$(X)$ $\geqslant$ \textit{minutil}, then $X$ passes the $RU$ strategy.
\end{property}
\textbf{Proof:} It is evident that for two itemsets $X$ and $Y$ where $X$ $\subseteq$ $Y$, only items $a$ $\in$ $I^{\star}$ that satisfy $X$ $\prec$ $a$ are extended onto $X$ to form $Y$. This implies that $U(Y)$ $<$ \textit{RU}$(X)$ + \textit{PU}$(X)$. Thus, we can have that $U(X)$ $<$ \textit{PU}$(X)$ $<$ \textit{RU}$(X)$ + \textit{PU}$(X)$.

\begin{property}[EUCP \cite{fournier2014fhm}]
    \rm  Let $m$ and $n$ be two distinct items. Utilizing the information in EUCS, if \textit{RTWU}$(m,n)$ $<$ \textit{minutil}, then all itemsets containing the itemset $\{m,n\}$ are not HUIs and can be pruned.
\end{property}
\textbf{Proof:} Based on Property \ref{RTWUPruning}, we can conclude that if \textit{RTWU}$(X)$ $<$ \textit{minutil}, then $X$ and its extensions exhibit low utility. We can confidently discard these itemsets as they will not affect the final results.

\begin{property}[LA pruning \cite{krishnamoorthy2015pruning}]
    \label{LAPruning}
    \rm Let $X$ and $Y$ are two different itemsets. If \textit{PU}$(X)$ + \textit{RU}$(X)$ - $\sum_{\forall T_k \in D \wedge  X \subseteq T_k \wedge  Y \nsubseteq T_k}$  \(\textit{PU}(X,T_k) + \textit{RU}(X,T_k)\) $<$ \textit{minutil}, then $XY$ is not a high-utility itemset, and neither are the extensions of $XY$. They can be pruned directly.
\end{property}

\textbf{Proof:} Let $X$ be a subset of $X'$ and $Y$ be a subset of $Y'$. We can derive the following formula: (1) \textit{PU}$(X)$ + \textit{RU}$(X)$ - $\sum_{\forall T_k \in D \wedge  X \subseteq T_k \wedge  Y \nsubseteq T_k}$  \(\textit{PU}(X,T_k) + \textit{RU}(X,T_k)\) = $\sum_{\forall T_k \in D}$  \(\textit{PU}(X,T_k) + \textit{RU}(X,T_k)\) - $\sum_{\forall T_k \in D \wedge  X \subseteq T_k \wedge  Y \nsubseteq T_k}$  \(\textit{PU}(X,T_k) + \textit{RU}(X,T_k)\) = $\sum_{\forall T_k \in D \wedge  X \subseteq T_k \wedge  Y \subseteq T_k}$  \(\textit{PU}(X,T_k) + \textit{RU}(X,T_k)\) + $\sum_{\forall T_k \in D \wedge  X \subseteq T_k \wedge  Y \nsubseteq T_k}$  \(\textit{PU}(X,T_k) + \textit{RU}(X,T_k)\) - $\sum_{\forall T_k \in D \wedge  X \subseteq T_k \wedge  Y \nsubseteq T_k}$  \(\textit{PU}(X,T_k) + \textit{RU}(X,T_k)\) = $\sum_{\forall T_k \in D \wedge  X \subseteq T_k \wedge  Y \subseteq T_k}$  \(\textit{PU}(X,T_k) + \textit{RU}(X,T_k)\). (2) \textit{U(X'Y')} = $\sum_{X'Y'\subseteq T_k\in D}$\(U(X'Y',T_k)\) $\le$ $\sum_{XY\subseteq T_k\in D}$ \(PU(XY,T_k) + RU(XY,T_k)\) $\le$ $\sum_{\forall T_k \in D \wedge  X \subseteq T_k \wedge  Y \subseteq T_k}$  \(\textit{PU}(XY,T_k) + \textit{RU}(XY,T_k)\) $\le$ $\sum_{\forall T_k \in D \wedge  X \subseteq T_k \wedge  Y \subseteq T_k}$  \(\textit{PU}(X,T_k) + \textit{RU}(X,T_k)\) $<$ \textit{minutil}. Based on the derivations of formulas (1) and (2) above, we can prove Property \ref{LAPruning}. More detailed proof can be referred to Ref. \cite{krishnamoorthy2015pruning}.

\begin{property}[maxPer pruning \cite{fournier2016phm}]
    \label{maxPer_Pruning}
    \rm For an itemset $X$, if \textit{maxPer}$(X)$ $>$ \textit{maxPer}, then $X$ and its super-itemsets do not qualify as periodic itemsets. Conversely, if \textit{maxPer}$(X)$ $\leqslant$ \textit{maxPer}, the potential periodicity of itemset $X$ and its extensions warrants further investigation.
\end{property}
\textbf{Proof:} This observation is self-evident. For any given itemsets $X$ and $Y$, with $Y$ being a super-itemset of $X$, it is apparent that $T(X)$ $\supseteq$ $T(Y)$. Consequently, the period of $X$ cannot exceed that of $Y$. Thus, we conclude that \textit{maxPer}$(Y)$ $\geq$ \textit{maxPer}$(X)$ $>$ \textit{maxPer}.
 
\begin{property}[maxAvg pruning \cite{fournier2016phm}]
\label{maxAvg_Pruning}
    \rm Given an itemset $X$, if \textit{avgPer}$(X)$ $>$ \textit{maxAvg}, then both itemset $X$ and its super-itemsets are not periodic itemsets. If \textit{avgPer}$(X)$ $\leqslant$ \textit{maxAvg}, then itemset $X$ and its extensions may potentially be periodic itemsets, requiring further investigation.
\end{property}
\textbf{Proof:} According to the formula in Ref. \cite{fournier2016phm}, \textit{avgPer}$(X)$ = $|\mathcal{D}|$ $/$ ( $|$ \textit{sup}$(X)$ $|$ $+$ 1), if \textit{avgPer}$(X)$ $>$ \textit{maxAvg}, which means \textit{avgPer}$(X)$ = $|\mathcal{D}|$ $/$ ( $|$ \textit{sup}$(X)$ $|$ + 1) $>$ \textit{maxAvg}, then \textit{sup}$(X)$ $<$ $|\mathcal{D}|$ $/$ \textit{maxAvg} - 1. For a given itemset $X$ and $Y$, where $Y$ is a superset of $X$, it is evident that \textit{sup}$(Y)$ $\leqslant$ \textit{sup}$(X)$ $<$  $|\mathcal{D}|$ $/$ \textit{maxAvg} - 1, so \textit{avgPer}$(Y)$ $>$ \textit{maxAvg}.

\subsection{Threshold-raising strategy}

Here, we introduce several threshold-raising strategies used in the TPU algorithm.

\begin{strategy}[Periodic real Item Utility strategy, PIU]
    \rm This is an improved version of the RIU threshold-raising strategy \cite{ryang2015REPT}. PIU also works after the initial database scan. During this scan, we compute utility and periodicity information for all single items. In $\mathcal{D}$, the utility of an item $i$ that satisfies the periodicity constraint is denoted as \textit{piu(i)}. Note that here we record only the utility of the items that satisfy the periodic constraints, not the utility of all items, which is a difference from the RIU threshold-raising strategy. The PIU strategy leverages the utility of single items with periodicity to raise the \textit{minutil} threshold.
\end{strategy}

Let $I$ = \{$i_1$, $i_2$, $i_3$, $\cdots$, $i_m$\} represent the set of items in $\mathcal{D}$, and let $P$ = \{$piu_1$, $piu_2$, $piu_3$, $\cdots$, $piu_n$\} represent the array of item utilities in $I$ that satisfy the periodic constraints, where $n$ $\leqslant$ $m$. Let $piu_k$ be the $k$-th highest value in \textit{P}. The PIU strategy raises the \textit{minutil} threshold to $piu_{k}$ (1 $\leqslant$ $k$ $\leqslant$ $n$). This new value serves as the \textit{minutil} during the mining process until it is further raised by other threshold-raising strategies. For example, consider Table \ref{Database} with \textit{minPer} = 1, \textit{maxPer} = 5, \textit{minAvg} = 1, \textit{maxAvg} = 5, and $k$ = 3. Since we need to discover the top-$k$ periodic itemsets in $\mathcal{D}$, if we obtain the $k$-th highest utility itemset (\textit{Part}$_k$) by observing a portion of the itemsets, then the utility of the top-$k$ itemsets in the dataset ($\textit{Data}_k$) must be greater than or equal to the utility value of \textit{Part}$_k$. After the first scan of $\mathcal{D}$, the utility and periodicity information for all single items in $\mathcal{D}$ will be determined, as presented in Table \ref{table:piu} and Table \ref{table:singleitemperiodic}. Note that the \textit{piu} and \textit{riu} values for negative items are not displayed here, as \textit{minutil} cannot be a negative number. The 3rd highest value in $P$ is 35, thus the \textit{minutil} can be raised to 35.

\begin{table}[h]
    \centering
    \caption{The RIU and PIU of the single item}
    \begin{tabular}{|c|c|c|c|c|c|c|c|c|}
    \hline
    \textbf{Item} & $g$ & $e$ & $c$ & $a$ & $d$ & $b$ & $f$\\ \hline
    \textbf{RIU}  & - & - & - & 20 & 54 & 35 & 132 \\ \hline
    \textbf{PIU}  & - & - & - & 20 & 54 & 35 & 132 \\ \hline
    \end{tabular}
    \label{table:piu}
\end{table}

\begin{strategy}
    \rm \textbf{Periodic Co-occurrence Utility Descending strategy, PCUD}. This novel strategy is used during the second database scan. PCUD leverages the utility of 2-itemsets recorded in the PEUCS to enhance the \textit{minutil} threshold. Using PEUCS can avoid numerous joint operations during utility list construction. PEUCS is designed to store and quickly traverse the \textit{RTWU}, \textit{utility}, and periodicity of item pairs. If the \textit{RTWU} of each entry recorded in the PEUCS for item pairs is not less than \textit{minutil} and satisfies the periodicity constraints, then the pair of items is a potential PHUI. Thus, the utility of 2-itemsets in the PEUCS can be used to raise the threshold. To illustrate this process more clearly, we introduce the PCUDM, which helps us to understand the specific procedure. PCUDM is a utility matrix that records the utility of itemsets in the PEUCS that meet the periodicity conditions and the \textit{RTWU} pruning strategy.
\end{strategy}

\begin{table}[h]
    \centering
    \caption{PCUDM}
    \begin{tabular}{|c|c|c|c|c|c|c|}
    \hline
    \textbf{Item} & $e$ & $c$ & $a$ & $d$ & $b$ & $f$\\ \hline
    $g$     & $-$ & $-$ & $\times$  & 20 &  8 & 20 \\ \hline
    $e$   &  & $-$ & $-$ & 4 & $\times$  & 44 \\ \hline
    $c$ &   & & $\times$ & 27 & 23 & 77 \\ \hline
    $a$    &  & & & 41 & $\times$ & $\times$ \\ \hline
    $d$     &  & & &  &  38 & 90 \\ \hline
    $b$  &  & & & &  &  120 \\ \hline
    \end{tabular}
    \label{table:PCUDM}
\end{table}

Let $P_{D}(k)$ refer to the top-$k$ PHUIs in $\mathcal{D}$ and $\textit{PCUD}_k$ denote the $k$-th highest utility in PCUDM. We have $U(X)$ $\geqslant$ $\textit{PCUD}_k$ for all $X$ $\in$ $P_{D}(k)$. By leveraging the information in the PCUDM, the PCUD threshold-raising strategy can elevate \textit{minutil} to the $k$-th highest utility in PCUDM. Considering Table \ref{Reordered_Database}, given $k$ = 3, \textit{minPer} = 1, \textit{maxPer} = 5, \textit{minAvg} = 1, \textit{maxAvg} = 5, and $k$ = 3, we can use Definition \ref{definition:ItemsetPeriod} to calculate that the support threshold is 1 $\leqslant$ \textit{sup}$(X)$ $\leqslant$ 9. By applying periodicity constraints and the \textit{RTWU} strategy to the data in Table \ref{table:PEUCS}, we can obtain the PCUDM, which is demonstrated in Table \ref{table:PCUDM}. The third-highest utility in PCUDM is 77; thus, the PCUD strategy can raise \textit{minutil} to 77. Note that in Table \ref{table:PCUDM}, the symbol $-$ indicates that the itemset's utility is less than 0, and $\times$ indicates that the itemset does not meet the periodicity conditions.

\begin{strategy}[Periodic Real Utility strategy, PRU]
    \rm This novel strategy is employed during the second scan of the database. It uses the utility of all single items and pair itemsets that satisfy periodicity conditions to raise the \textit{minutil} threshold. Here, a priority queue $Q$ is used to maintain the utility of the currently mined top-$k$ PHUIs. Let $P_{D} (k)$ denote the top-$k$ PHUIs in $\mathcal{D}$. Let $Q_{D}(k)$ symbolize the $k$-th largest utility in $Q$. We have $U(X)$ $\geqslant$ $Q_{k}$, $\forall$ $X$ $\in$ $P_{D}(k)$. Using the information in $Q$, the PRU threshold-raising strategy can raise \textit{minutil} to the $k$-th largest utility in $Q$.
\end{strategy}

In Table \ref{Database} and setting \textit{minPer} = 1, \textit{maxPer} = 5, \textit{minAvg} = 1, \textit{maxAvg} = 5, and \textit{k} = 3, the information for all single items and pair itemsets that meet the periodicity conditions is stored in Table \ref{table:piu} and Table \ref{table:PCUDM}, respectively. From Table \ref{table:piu} and Table \ref{table:PCUDM}, we can observe that the top-3 itemsets in terms of utility are \{$f$:132\}, \{$bf$:120\}, and \{$df$:90\}. Therefore, the PRU strategy can raise the \textit{minutil} to 90.

\subsection{The TPU algorithm}

Here, we introduce the TPU algorithm, which is divided into two principal phases: the preparation phase and the search phase. The pseudocode of the preparation phase is shown in Algorithm \ref{algo:TPU}, the pseudocode of the search phase is shown in Algorithm \ref{algo:Search}, and the construction of the utility-list during the search phase is detailed in Algorithm \ref{algo:Construct}.

\begin{algorithm}[ht]
    \small
    \caption{The TPU algorithm}
    \label{algo:TPU}
    \LinesNumbered
    \KwIn{$\mathcal{D}$: the database; $k$: the desired count of PHUIs by the user; \textit{minutil}: the minimum utility of top-$k$ HUIs found so far; \textit{maxPer}, \textit{minPer}, \textit{maxAvg}, \textit{minAvg}: the periodic constraints.}
    \KwOut{a set of top-$k$ PHUIs.}
    
    scan $\mathcal{D}$ and compute the \textit{TWU} and periodic information \textit{maxPer}, \textit{minPer}, \textit{avgPer} for all individual items\;
    apply the \textit{PIU} strategy to raise \textit{minutil}\;
    $I^{\star}$ $\leftarrow $ each item $i$, which makes \textit{TWU}$(i)$ $\geqslant $ \textit{minutil}, \textit{maxPer}$(i)$ $\leq$ \textit{maxPer} and \textit{avgPer}$(i)$ $\leqslant$ \textit{maxAvg}\;
    order the items in $I^{\star}$ according to the $\prec$ rule\;
    scan $\mathcal{D}$ again and construct the \textit{PNUL} and \textit{PEUCS}\;
    apply the \textit{PCUD} and \textit{PRU} strategies to increase the \textit{ minutil}\;
    call \textit{Search}(\o, \textit{null}, \textit{PNULs}, \textit{minutil}, \textit{maxPer}, \textit{minPer}, \textit{maxAvg}, \textit{minAvg})\;
    \Return a set of top-$k$ PHUIs
\end{algorithm}

\begin{algorithm}[ht]
    \small
    \caption{The Search procedure}
    \label{algo:Search}
    \LinesNumbered
    \KwIn{$P$: a prefix of a pattern; \textit{P-UL}: the utility-list of $P$; \textit{ULs}: the utility-lists of all extensions of $P$; \textit{minutil}, \textit{minPer}, \textit{maxPer}, \textit{minAvg}, and \textit{maxAvg}.}
    \KwOut{the top-$k$ PHUIs.}
    
    \For{\rm each $P_{m}$ $\in$ \textit{ULs}}{
        \If{\rm  $P_{m}$ satisfies the periodicity and utility constraints}{
            top-$k$ PHUIs $\leftarrow$ top-$k$ PHUIs $\cup$ $P_{m}$\;
            \textit{minutil} = \textit{min}\{$U(P_{m})$ $\mid$ $P_{m}$ $\in$ \textit{top-$k$ PHUIs}\}\;
        }
        \If{\textit{PU}($P_{m}$) + \textit{RU}($P_{m}$) $\geq$ \textit{minutil} \&\& \textit{maxPer}($P_{m}$) $\leq$ \textit{maxPer} \&\& \textit{avgPer}($P_{m}$) $\leq$ \textit{maxAvg}}{
            $P_{m}$.\textit{ULs} $\leftarrow$ $\varnothing$\;
            \For{\rm  each $P_{n}$ $\in$ \textit{ULs} \&\& $m$ $\prec$ $n$}{
                \If{\rm items $m$ and $n$ satisfy the \textit{EUCP} \&\& \textit{ESCP}}{
                    $P_{mn}$ = $P_{m}$ $\cup$ $P_{n}$\;
                    $P_{mn}$.\textit{UL} $\leftarrow$ \textit{Construct}($P$, $P_{m}$, $P_{n}$)\;
                    $P_{m}$.\textit{ULs} $\leftarrow$ $P_{m}$.\textit{ULs} $\cup$ $P_{mn}$\;
                }
            }
            call \textit{Search}($P_{m}$, $P_{m}$.\textit{UL}, $P_{m}$.\textit{ULs}, \textit{minutil}, \textit{minPer}, \textit{maxPer}, \textit{minAvg}, \textit{maxAvg})\;
        }
    }
    \Return the top-$k$ PHUIs
\end{algorithm}

\begin{algorithm}[ht]
    \small
    \caption{The Construct procedure}
    \label{algo:Construct}
    \LinesNumbered
    \KwIn{$P$: an itemset with its \textit{PNU-list}; $P_m$: the extension of $P$ with an item $m$; $P_n$: the extension of $P$ with an item $n$.}
    \KwOut{$P_{mn}$.}
    
    initialize $P_{mn}$.\textit{UL} $\leftarrow$ $\emptyset$, \textit{sumUtility} = $PU(P_{m})$ + $RU(P_{m})$\;

    \For{\rm tuple $em$ $\in$ $P_{m}$.\textit{UL}}{
        \eIf{$\exists$ $en$ $\in$ $P_{n}$.\textit{UL} \&\& \textit{em.tid} == \textit{en.tid}}{
            \eIf{$P$.\textit{UL} == \textit{null}}{
                call \textit{getPeriod}(\textit{em.tid}, $P_{mn}$.\textit{UL})\;
                \textit{emn} $\leftarrow$ \textit{[em.tid}, \textit{em.pu} $+$ \textit{en.pu}, \textit{em.nu} $+$ \textit{en.nu}, \textit{en.ru]}\;
            }{
                search $e$ $\in$ $P$.\textit{UL} \&\& \textit{e.tid} == \textit{em.tid}\;
                \textit{getPeriod}(\textit{em.tid}, $P_{mn}$.\textit{UL})\;
                \textit{emn} $\leftarrow$ \textit{[em.tid}, \textit{em.pu} $+$ \textit{en.pu} $-$ \textit{e.pu}, \textit{em.nu} $+$ \textit{en.nu} $-$ \textit{e.nu}, \textit{en.ru]}\;
            }
            $P_{mn}$.\textit{UL} $\leftarrow$ $P_{mn}$.\textit{UL} $\cup$ \textit{emn}\;
        }{
            \textit{sumUtility} $-=$ (\textit{em.pu} + \textit{em.ru})\;
            \textit{sup}$(P_{m})$$--$\;
            \If{\textit{sumUtility} $<$ \textit{minutil} $\parallel$ \textit{sup}($P_{m}$) $<$ \textit{supportPruning}}{\Return \textit{null}\;}
        }
    }
    call \textit{getPeriod}(\textit{em.tid}, $P_{mn}$.\textit{UL})\;
    \Return $P_{mn}$.\textit{UL}
\end{algorithm}

\begin{algorithm}[ht]
    \small
    \caption{getPeriod}
    \label{algoGetPeriod}
    \LinesNumbered
    \KwIn{\textit{em.tid}, $P_{mn}$.\textit{UL}.}
    \KwOut{\textit{newPeriod}.}
    
    $\textit{period}_{emn}$ = \textit{calculatePeriod}(\textit{em.tid}, $P_{mn}$.\textit{UL})\;
    \If{$\textit{period}_{emn}$ $>$ \textit{maxPer}}{\Return \textit{null}\;}
    \textit{newPeriod} $\leftarrow$ \textit{UpdatePeriodInformation}($P_{mn}$.\textit{UL}, $\textit{period}_{emn}$)\;

    \Return \textit{newPeriod}
\end{algorithm}

\textbf{Preparatory phase}. In this paper, a set-enumeration tree \cite{liu2012mining} can visualize the search space for the TPUM problem. Exploration of this structure is conducted via a depth-first search approach. The order of item processing is as defined in Definition \ref{OrderofItems}. The initial value of \textit{minutil} is 0. It is continuously raised during the subsequent execution of the algorithm. The pseudocode for the preparation phase of the TPU algorithm is shown in Algorithm \ref{algo:TPU}. Our algorithm scans the database twice. During the first scan, it calculates the \textit{RTWU} and periodic information (\textit{maxPer}, \textit{minPer}, and \textit{support}) for single items (line 1). Then, it raises the threshold using the \textit{PIU} strategy (line 2). Next, it filters out unpromising items using strategies Property \ref{RTWUPruning}, Property \ref{maxAvg_Pruning}, and Property \ref{maxPer_Pruning} (line 3). The items that meet the pruning criteria are recorded in $I^{\star}$ and sorted according to Definition \ref{OrderofItems} (line 4). During the second scan, the database is reordered, and the \textit{PNUL} and \textit{PEUCS} structures for single items are calculated (line 5). Then, it raises the threshold using the \textit{PCUD} and \textit{PRU} strategy (Line 6). Algorithm \ref{algo:TPU} finally calls Algorithm \ref{algo:Search} to commence the actual mining process (line 7). Setting \textit{minPer} = 1, \textit{maxPer} = 5, \textit{minAvg} = 1, \textit{maxAvg} = 5, and $k$ = 3. First, the \textit{PIU} strategy raises \textit{minutil} to 35. All single items satisfy strategies Property \ref{RTWUPruning}, Property \ref{maxAvg_Pruning}, and Property \ref{maxPer_Pruning}, thus no single items are deleted. The order of items is \{$g$ $\prec$ $e$ $\prec$ $c$ $\prec$ $a$ $\prec$ $d$ $\prec$ $b$ $\prec$ $f$\}. The reordered database and the \textit{PEUCS} structure are shown in Tables \ref{Reordered_Database} and \ref{table:PEUCS}, respectively. The \textit{PCUD} strategy raises \textit{minutil} to 77. The \textit{PRU} strategy raises \textit{minutil} to 90.

\textbf{Search phase}. The pseudocode for the search stage of TPU, as shown in Algorithm \ref{algo:Search}, has three input parameters. $P_m$ is a prefix itemset of length $l$. $P_{m}$.\textit{UL} is the UL of $P_{m}$, and $P_{m}$.\textit{ULs} is the UL of extensions of $P_{m}$. This algorithm is recursive. Firstly, it checks if $P_{m}$ is a potential periodic high utility itemset (line 2). If so, $P_{m}$ is added to the \textit{top-$k$ PHUIs} and updates \textit{minutil} (lines 3 to 4). Then, if $P_{m}$ satisfies Property \ref{RU_Pruning}, Property \ref{maxAvg_Pruning}, and Property \ref{maxPer_Pruning}, we consider $P_{m}$ as a new prefix and calculate its extensions $P_{m}$.\textit{ULs} (lines 6 to 14). Finally, the TPU algorithm recursively executes \textit{Search}($P_{m}$, $P_{m}$.\textit{UL}, $P_{m}$.\textit{ULs}) with the new parameters (line 15). The TPU algorithm, initially, is invoked with parameters $ \varnothing $, the set of items \{$g$, $e$, $c$, $a$, $d$, $b$, $f$\}, and their utility lists as inputs. For item $ g $, $U(g)$ < 0 < 90, indicating it's not a PHUI. It checks whether $g$ can be extended. Since $PU(g)$ + $RU(g)$ = 88 < 90, it does not satisfy Property \ref{RU_Pruning}, and thus $ g $ cannot be extended. Next, we evaluate the item $ e $. $U(e)$ < 0 < 90 suggests that $ e $ is not a PHUI. However, as $PU(e)$ + $RU(e)$ = 118 > 90, \textit{maxPer}$(e)$ = 3 < 5, and \textit{avgPer}$(e)$ = 1.43 < 5, $e$ can be extended. Using the information from \textit{PEUCS} in Table \ref{table:PEUCS}, we obtain the \textit{RTWU} and periodicity information for \{$ec$, $ea$, $ed$, $eb$, $ef$\}. Among these, only \textit{RTWU}$(ea)$ > 90 satisfies Property \ref{RTWUPruning}. Hence, only the utility list for $ea$ should be constructed. This process is repeated with $ea$ as the prefix. In total, apart from single items, TPU needs to construct the utility lists for \textit{ea}, \textit{cd}, \textit{cdf}, \textit{cb}, \textit{cbf}, \textit{cf}, \textit{df}, and \textit{bf}.

\textbf{Construct procedure}. The pseudocode for constructing the UL is demonstrated in Algorithm \ref{algo:Construct}. The construction is similar to that in PHM \cite{fournier2016phm} and PHMN \cite{lai2023PHMN}. The inputs are the \textit{PNUL} of $P_m$, $P_n$, and its prefix $P$, and the output is the \textit{PNUL} of $P_{mn}$. It starts by initializing $P_{mn}$.\textit{UL} and \textit{sumUtility} (line 1). Then, it iterates through each \textit{tuple}, $em$, in $P_{m}$.\textit{UL} (line 2). If there is no \textit{tuple}, \textit{en}, in $P_{n}$.\textit{UL} with the same \textit{tid}, it updates \textit{sumUtility} and \textit{sup}$(P_{m})$, and determines whether to prune (lines 14 to 18). Otherwise, it updates the periodic information and $P_{mn}$.\textit{UL} based on whether $P$.\textit{UL} is \textit{null} (lines 4 to 12). Finally, it computes the last period and decides whether to construct $P_{mn}$.\textit{UL} (line 21).

\subsection{Complexity analysis}

The TPU algorithm is mainly divided into two phases: the preparation phase and the search phase. The search phase includes the construction phase of the utility list. Let us assume that $\mathcal{D}$ contains $m$ transactions and $n$ items, and the average length of the transactions is $a$. The algorithm scans $\mathcal{D}$ and calculates each item's TWU and periodicity information. In this step, the time and space complexity are $O$($am$) and $O$$(n)$, respectively. The time complexity and space complexity of applying the PIU strategy are $O$$(n)$ and $O$$(1)$, respectively. We assume that there are $b$ items that satisfy the pruning strategies of TWU, maxPer, and maxAvg. Then, the time complexity and the space complexity of constructing $I^{\star}$ are $O(n)$ and $O(b)$, respectively. In the second scan of $\mathcal{D}$, the construction of PNUL and PEUCS incurs a time complexity of $O$$(am)$ and a space complexity of $O$$(b$ + $b^2)$. The application of the PCUD and PRU strategies has a time complexity of $O$$(b^2)$ and a space complexity of $O(1)$. Therefore, the overall time complexity and the space complexity of the preparation phase are $O$$(2am$ + $2n$ + $b^2)$ and $O$($n$ + $b^2$ + $2b$ + $2)$, respectively. The search phase follows a recursive approach, where the complexity is primarily determined by the following seven factors: the number of iterations, the recursion depth, and the computational complexity of operations within each iteration and recursive call. Initially, the size of $ULs$ corresponds to the number of single items that satisfy the given conditions, denoted as $b$. The maximum itemset length is $b$, leading to a recursion depth of at most $b$ layers. Let $O(T)$ and $O(S)$ denote the time and space complexity of each invocation of the Construct algorithm, respectively. In the worst case, the time complexity and space complexity of the search algorithm are $O$$(n^b$ $\times$ $T)$ and $O$$(n^b$ $\times$ $S$ + $k)$, respectively. Thus, the overall time complexity and space complexity of the TPU algorithm are $O$$(n^b$ $\times$ $T$ + $2am$ + $2n$ + $b^{2})$ and $O$($n^{b}$ $\times$ $S$ + $k$ + $n$ + $b^{2}$ + $2b$ + 2), respectively.

\section{Experiments} \label{sec: experiments}

This section details the comprehensive experiments performed to verify the correctness and evaluate the performance of the TPU algorithm. TPU is the first algorithm designed to address the TPUM problem that involves both positive and negative utility values. To verify the correctness of the algorithm, we compared the number of PHUIs discovered by the TPU with those discovered by the PHMN algorithm \cite{lai2023PHMN}. The existing top-$k$ HUIM algorithms cannot be applied to this problem because they do not consider itemset periodicity. Therefore, in terms of effectiveness and efficiency evaluation, existing top-$k$ HUIM algorithms cannot be compared with the TPU proposed in this paper. To assess TPU's efficiency, we compared two variants. Each employs different threshold-raising strategies: TPU$_{\rm base}$ uses only the PIU strategy; TPU utilizes the PIU, PCUD, and PRU strategies. The algorithms were developed using Java, and experiments were carried out on a DELL workstation featuring a 12th Gen Intel(R) Core(TM) i7 processor, Windows 11, and 16 GB of RAM.

\subsection{Datasets}

The experiment uses a total of six datasets, which are Chess\_negative, accidents\_negative, mushroom\_negative, kosarak\_negative, retail\_negative, and pumsb\_negative. These datasets are available for download from the open-source data mining library at \href{https://www.philippe-fournier-viger.com/spmf/}{https://www.philippe-fournier-viger.com/spmf/.} Table \ref{CharacteristicsOfDataset} provides detailed characteristics of the datasets, where \textit{NT} denotes the number of transactions, \textit{NI} indicates the number of items, and \textit{ATL} stands for the average transaction length. The \textit{Periodic threshold} represents the periodicity threshold set for testing each dataset. The first number denotes \textit{minPer}, the second denotes \textit{maxPer}, the third denotes \textit{minAvg}, and the fourth denotes \textit{maxAvg}. The distribution of items in the dataset and the dataset's density affect the setting of the periodicity threshold. If the items in the dataset are sparsely distributed, the number of itemsets that satisfy the periodicity constraints will be smaller, requiring a more relaxed range for the periodicity constraints. On the other hand, if the dataset has a high density, meaning the items are relatively concentrated, there will be more itemsets satisfying the periodicity constraints, and thus the range of the periodicity constraints should be more stringent. The periodicity in the datasets used in this study is based on the settings in Ref. \cite{fournier2016phm} and Ref. \cite{lai2023PHMN}, while also considering the specific requirements when mining for top-$k$ patterns. These datasets are real; however, their external utilities and quantities are generated using lognormal and uniform distributions, respectively. These distributions are set up similarly to the settings of previous studies \cite{gan2020tophui,lai2023PHMN,lin2016fhn}. An overview of the datasets used in this study is provided below. Chess\_negative is derived from the UCI chess dataset. Accidents\_negative contains anonymized data from traffic accidents. Mushroom\_negative is based on the UCI mushrooms dataset. Kosarak\_negative consists of click-stream transaction data from a Hungarian news portal. Retail\_negative comprises customer transactions from an anonymous Belgian retail store. Pumsb\_negative contains census information related to population and housing. Each dataset's parameters $k$ were set to 10, 20, 30, 40, 50, 60, 70. Because each dataset has different characteristics, we use different period thresholds for different datasets. For example, on the pumsb\_negative dataset, when \textit{minPer} = 1, \textit{maxPer} = 20, \textit{minAvg} = 1, \textit{maxAvg} = 10, and $k$ = 10, the running time of the TPU algorithm for mining the top-10 periodic itemsets is 4200.79 seconds. If the periodicity threshold is set too broadly, it will result in a significantly prolonged runtime. This does not affect the fairness of the comparison test.

\begin{table*}[h]
    \centering
    \caption{Characteristics of the dataset}
    \label{CharacteristicsOfDataset}
    \begin{tabular}{|c|c|c|c|c|}
        \hline
        \textbf{Dataset} & \textbf{\textit{|NT|}} & \textbf{\textit{|NI|}} & \textbf{ATL} & \textbf{Periodic threshold} \\
        \hline
        Chess\_negative & 3196 & 75 & 37.0 & 1-50-1-20 \\
        \hline
        accidents\_negative & 340183 & 468 & 33.81 & 1-50-1-20\\
        \hline        
        mushroom\_negative & 8124 & 119 & 23.0 & 1-5000-1-2000\\
        \hline         
        kosarak\_negative & 9900022 & 41270 & 8.10 & 1-5000-1-2000\\
        \hline
        retail\_negative & 88162 & 16470 & 10.31 & 1-5000-5-500 \\ \hline
        pumsb\_negative & 49046 & 2113 & 74.0 & 1-10-1-10 \\  \hline
    \end{tabular}
\end{table*}

\subsection{Correctness analysis}

\begin{figure*}[h!]
  \centering
  \includegraphics[width=0.9\textwidth]{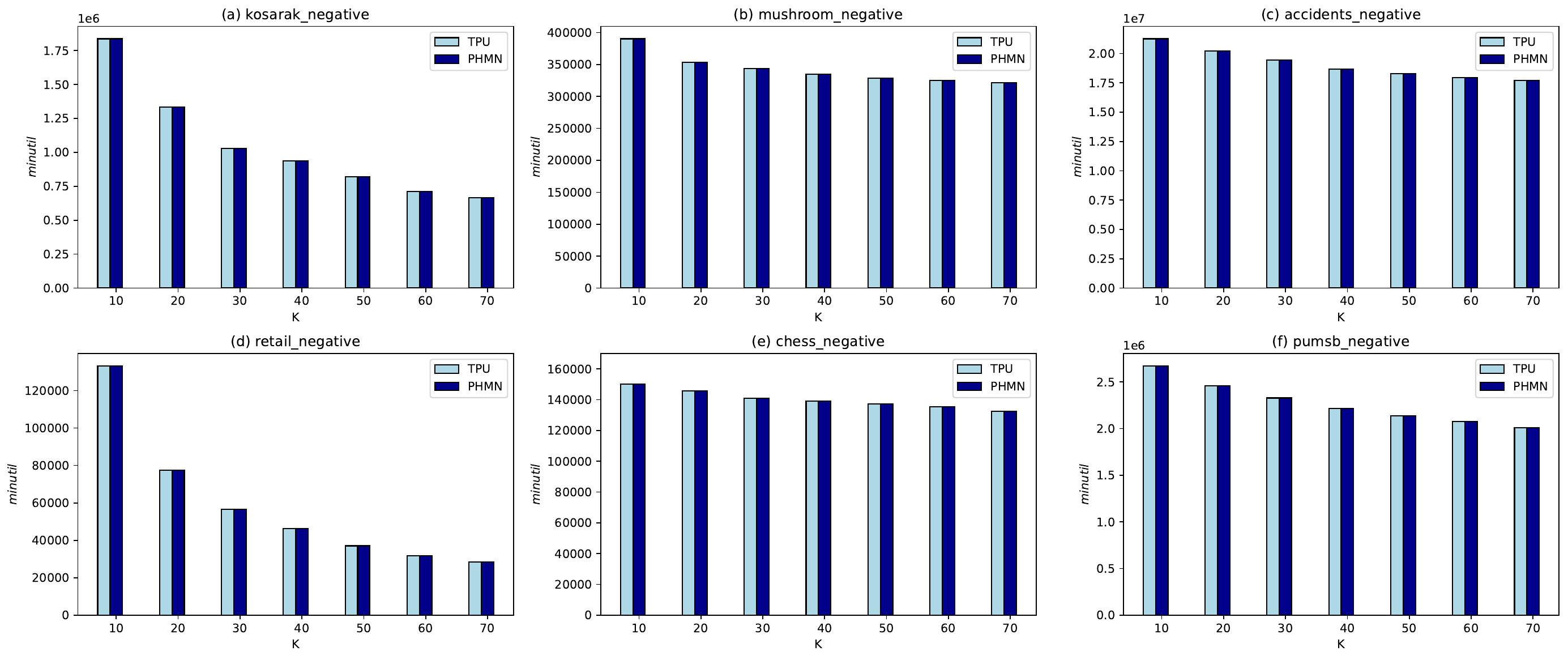} 
  \caption{\textit{minutil} under the varied top-$k$ value.}
  \label{fig:minutil}
\end{figure*}

To verify the correctness of the algorithm, we conducted a comparative analysis of the number of PHUIs identified by the TPU algorithm and those discovered by the PHMN algorithm \cite{lai2023PHMN}. Initially, we set the periodicity thresholds and the value of $k$. Using the TPU algorithm, we can discover the top-$k$ PHUIs and record the \textit{minutil} of the discovered top-$k$ PHUIs. Subsequently, we employed this \textit{minutil} along with the same periodicity thresholds to perform the mining task using the PHMN algorithm, documenting the number of PHUIs uncovered. Finally, we conducted a comparison of the number of PHUIs mined by both algorithms to ensure they were identical. As illustrated in Figure \ref{fig:minutil}, the experimental results confirm that the number of PHUIs identified by our algorithm is consistent with the number found by the PHMN algorithm, thereby validating the correctness of our proposed algorithm. Additionally, our findings reveal that the \textit{minutil} value decreases progressively as the value of $k$ increases.

\subsection{Candidate patterns}

\begin{figure*}[h!]
  \centering
  \includegraphics[width=0.9\textwidth]{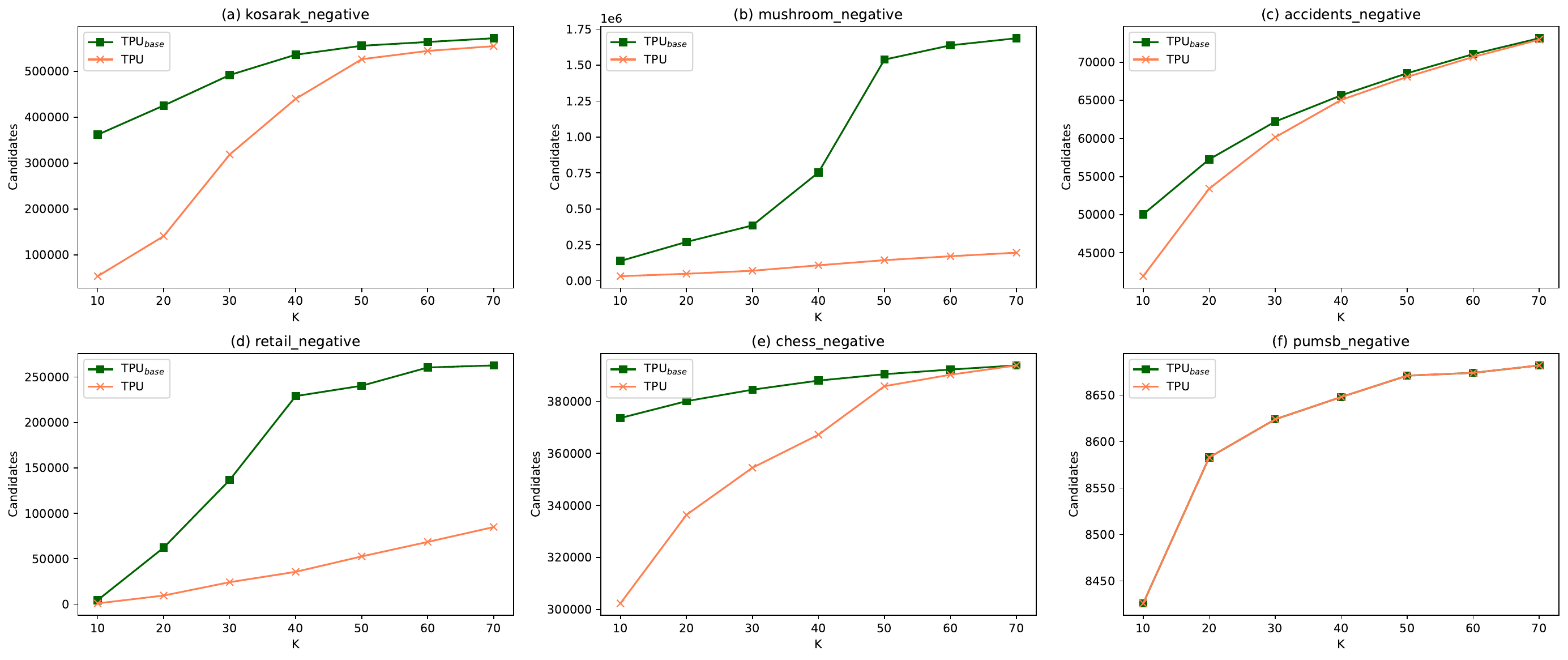} 
  \caption{Candidates under the varied top-$k$ value.}
  \label{fig:Candidate}
\end{figure*}

To assess the effectiveness of the TPU algorithm, we first compared the quantity of candidates generated by its two variants. For an equitable assessment, each test uses the same top-$k$ setting and records its generated candidates. Figure \ref{fig:Candidate} illustrates the number of candidates generated as the value of $k$ changes, given specific periodicity thresholds. It can be observed that, in the first five datasets, the TPU algorithm generates fewer candidates compared to TPU$_{\rm base}$. TPU employs all the threshold-raising strategies, whereas TPU$_{\rm base}$ utilizes only the PIU strategy. Therefore, the proposed PCUD and PRU strategies effectively raise the thresholds, allowing for the rapid elimination of unpromising patterns. However, as shown in Figure \ref{fig:Candidate}(f), for the pumsb\_negative dataset, the amount of candidates generated by both the TPU and TPU$_{\rm base}$ algorithms is identical. This is because, in this dataset, the number of 1-itemsets and 2-itemsets that satisfy the periodicity constraint is small. Moreover, the threshold-raising strategy only considers two items, and the utility of the threshold raised by the algorithm is too small compared to the utility of the final mined itemsets. In other words, our strategy cannot effectively raise the threshold in the pumsb\_negative dataset, which leads to the TPU and TPU$_{\rm base}$ algorithms having the same candidate itemsets. In conclusion, the proposed strategies generally lower the volume of candidate itemsets and the search space in most cases. 

\subsection{Runtime analysis}

\begin{figure*}[h!]
  \centering
  \includegraphics[width=0.9\textwidth]{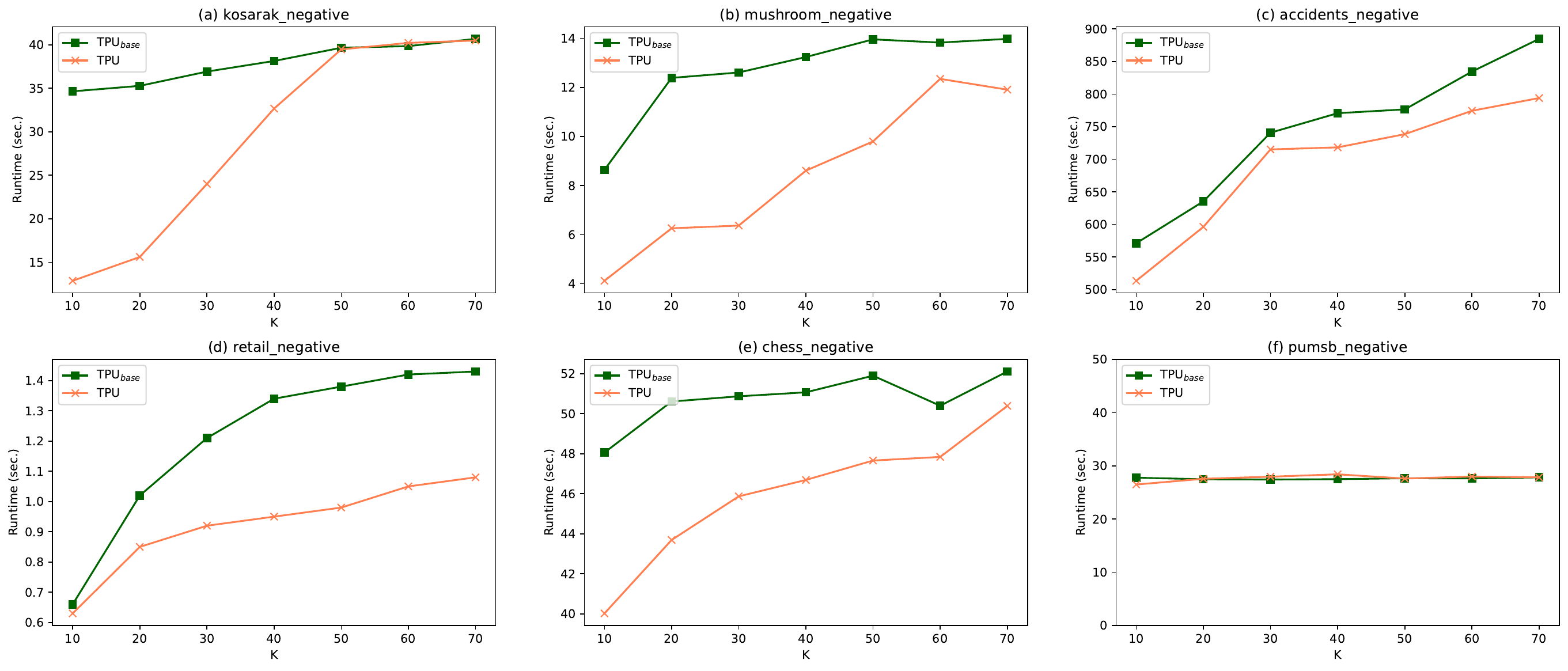} 
  \caption{Runtime under the varied top-$k$ value.}
  \label{fig:Runtime}
\end{figure*}

Here, the performance in terms of runtime is evaluated for both the TPU and TPU$_{\rm base}$ algorithms. Figure \ref{fig:Runtime} illustrates how the runtime varies with changes in the value of $k$, given specific periodicity thresholds. As the value of $k$ rises, it becomes apparent that the execution time of both algorithms shows an upward trend and TPU consistently exhibits shorter execution times compared to the TPU$_{\rm base}$ algorithm. This suggests that the strategies employed in TPU can expedite the execution process. For example, in the mushroom\_negative dataset, the Figure \ref{fig:Candidate}(b) shows that the number of candidate itemsets increases as the $k$ increases (indicating an expanding search space), and the Figure \ref{fig:Runtime}(b) shows that the runtime also increases with the $k$. This happens because as the $k$ increases, the number of PHUIs that need to be mined increases, and as a result, the uplifted utility threshold decreases. This leads to a larger number of candidate itemsets as the $k$ increases, which in turn causes the runtime to increase. However, as shown in Figure \ref{fig:Runtime}(f), the performance of the algorithms on the pumsb\_negative dataset is less optimal. Initially, TPU's execution time is shorter than that of the TPU$_{\rm base}$ algorithm. When $k$ grows, TPU's execution time surpasses that of the TPU$_{\rm base}$ algorithm. For instance, when $k$ = 30, the execution time of TPU is 27.43 seconds, whereas TPU$_{\rm base}$ consumes 27.96 seconds. When $k$ = 40, the execution time of TPU is 27.50 seconds, compared to 28.41 seconds of TPU$_{\rm base}$. This phenomenon can be attributed to the fact that, in the pumsb\_negative dataset, the number of 1-itemsets and 2-itemsets that satisfy the periodicity constraint is limited. Additionally, the threshold-raising strategy only considers two items, which results in a relatively small increase in the threshold compared to the utility values of the mined itemsets. Consequently, the strategy cannot effectively raise the \textit{minutil} threshold, leading to identical candidate itemsets in both the TPU and TPU$_{\rm base}$ algorithms. This inefficiency in raising the \textit{minutil} threshold is more pronounced in the TPU algorithm, although the PCUD and PRU strategies introduce additional computational overhead. As shown in Figure \ref{fig:Candidate}(f), the TPU and TPU$_{\rm base}$ algorithms result in the same number of candidate itemsets, and TPU does not significantly improve the \textit{minutil} compared to TPU$_{\rm base}$. As a result, the added time from the PCUD and PRU strategies causes TPU to have slightly longer execution times as $k$ increases. However, the difference in execution times is minimal. For instance, when $k$ = 40, the largest time difference is only 0.91 seconds, and the second-largest difference, at $k$ = 30, is just 0.53 seconds, as shown in Figure \ref{fig:Runtime}(f).

\subsection{Memory cost analysis}

\begin{figure*}[h!]
  \centering
  \includegraphics[width=0.9\textwidth]{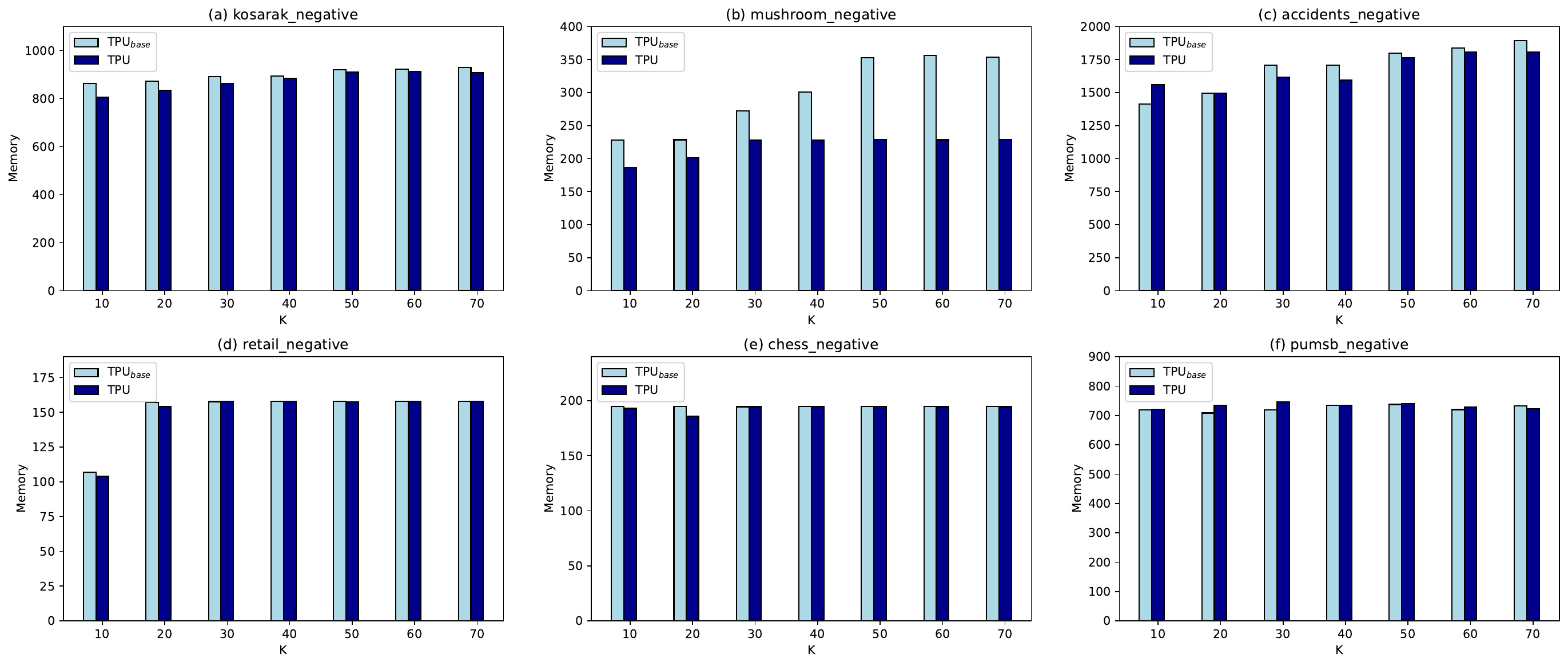} 
  \caption{Memory under the varied top-$k$ value.}
  \label{fig:Memory}
\end{figure*}

Figure \ref{fig:Memory} shows the memory consumption of both TPU and TPU$_{\rm base}$ algorithms. We can observe that TPU consumes a similar amount of memory as TPU$_{\rm base}$. However, on the mushroom\_negative dataset, TPU consumes significantly less memory than the TPU$_{\rm base}$  algorithm. As demonstrated in Figure \ref{fig:Memory}(b), as the value of $k$ increases from 10 to 70, the memory consumption of TPU$_{\rm base}$ is 228.33 MB, 228.61 MB, 272.57 MB, 300.73 MB, 352.59 MB, 355.97 MB, and 353.97 MB, respectively. In contrast, the memory consumption of TPU is 186.84 MB, 201.12 MB, 227.83 MB, 228.37 MB, 228.78 MB, 228.58 MB, and 228.77 MB, respectively. This is because the mushroom\_negative dataset is dense, with a larger number of 1-itemsets and 2-itemsets that satisfy the periodicity constraint, and the itemsets are more highly correlated. As a result, the TPU algorithm can quickly raise the threshold and prune unpromising itemsets, thereby reducing the cost of storing itemset information and leading to lower memory usage. This demonstrates that our proposed threshold-raising strategy effectively increases the threshold and speeds up the algorithm's execution without consuming excessive memory. Furthermore, the reduced memory usage on specific datasets, such as mushroom\_negative, indicates the efficiency of the TPU algorithm in handling large-scale data while maintaining optimal resource utilization. This efficiency is particularly beneficial in practical applications where both speed and memory usage are critical factors.

\subsection{Threshold-raising strategies}

This section presents the results of experiments conducted to evaluate the effectiveness of the threshold-raising strategies utilized in the algorithm. During the experiments, TPU utilized the PIU, PCUD, and PRU strategies for mining top-$k$ periodic high-utility itemsets with both positive and negative values, while TPU$_{\rm base}$ employed only the PIU strategy. By comparing the \textit{minutil} values achieved using the PIU, PCUD, and PRU strategies during the mining process, the six datasets outlined earlier will be used to assess the effectiveness of the threshold-raising strategies.

\begin{figure*}[h!]
  \centering
  \includegraphics[width=0.9\textwidth]{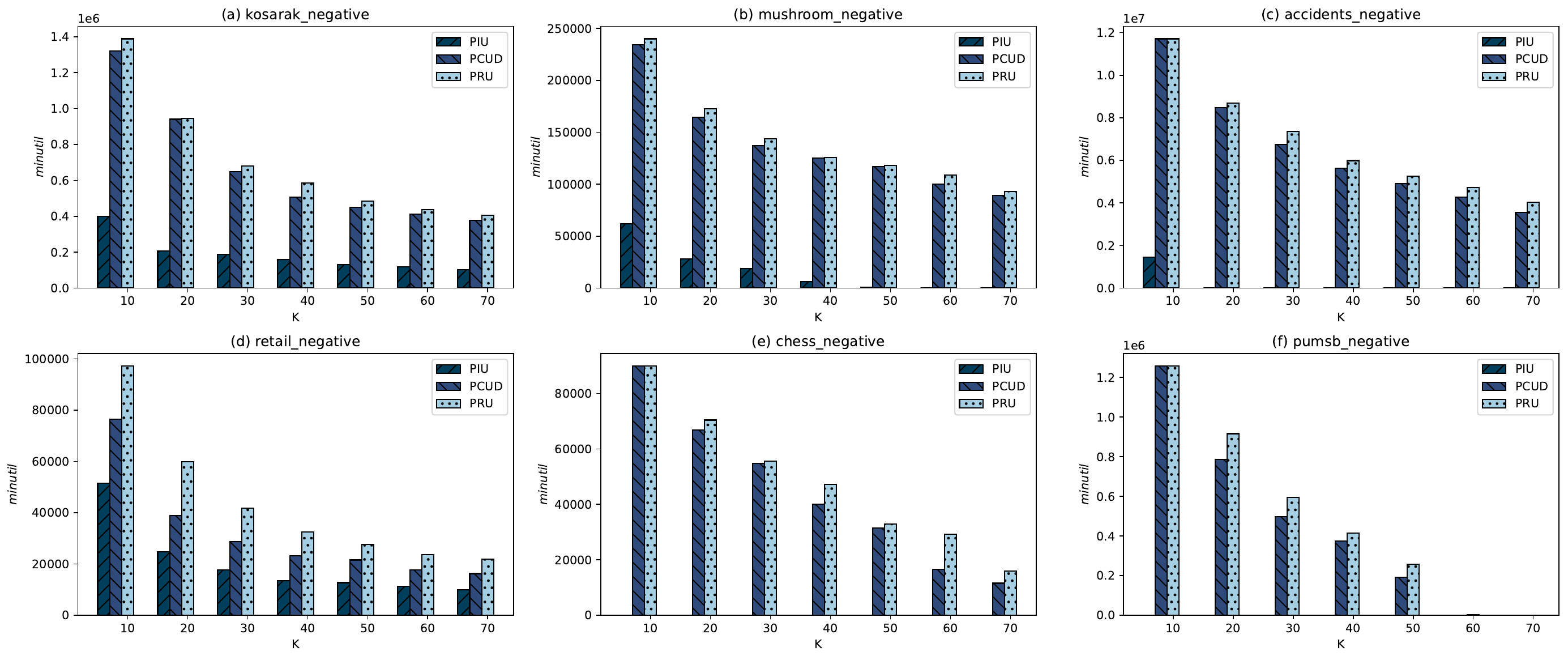}
  \caption{\textit{minutil} raising under the varied top-$k$ value.}
  \label{fig:strategy}
\end{figure*}

Figure \ref{fig:strategy} presents the experimental results, demonstrating that compared to the PIU strategy, both the PCUD and PRU strategies are more effective in enhancing the \textit{minutil}. The PRU strategy offers a modest improvement over the PCUD strategy in enhancing the \textit{minutil}. Overall, both the PCUD and PRU strategies are more effective than the PIU strategy in increasing the \textit{minutil}. Figure \ref{fig:strategy}(b) and Figure \ref{fig:strategy}(c) show that as $k$ increases, the column for the PIU strategy is not displayed. Similarly, Figure \ref{fig:strategy}(e) and Figure \ref{fig:strategy}(f) reveal that the PIU column is missing, with PIU values being zero. This is because, under the specific periodicity threshold setting, the number of 1-itemsets that satisfy the periodicity constraint in these datasets is less than $k$, so the PIU strategy cannot raise the threshold. Moreover, Figure \ref{fig:strategy}(f) indicates that when $k$ is 60 or 70, no data is displayed. This suggests that neither the PIU, PCUD, nor the PRU strategies can elevate the threshold under these conditions. The absence of data is due to the dataset containing fewer than $k$ single items and 2-itemsets that satisfy the periodic threshold. Therefore, when $k$ is large, the proposed threshold-raising strategies are ineffective in the pumsb\_negative dataset.

\section{Conclusion}  \label{sec: conclusion}

This paper introduces the TPU algorithm, which can successfully identify the top-$k$ PHUIs, including both positive and negative utilities. To enhance efficiency, we develop a novel data structure and three innovative threshold-raising strategies: PIU, PCUD, and PRU. Utilizing the information stored in PCUDM, we can swiftly determine whether to construct the UL of an itemset, thereby avoiding numerous UL join operations. The PIU, PCUD, and PRU strategies enable threshold raising. The proposed threshold-raising strategy achieves a runtime reduction of over 5\% on the datasets. The mushroom\_negative and kosarak\_negative datasets saw up to a 50\% reduction in runtime, while the chess\_negative dataset experienced a reduction of around 10\%. Additionally, memory consumption decreased by approximately 2\%, with the mushroom\_negative dataset showing the highest reduction of about 30\%. The experiments suggest that the TPU algorithm correctly discovers the top-$k$ PHUIs and the proposed threshold-raising strategies exhibit excellent performance in both efficiency and effectiveness. In future research, we plan to apply the proposed strategies to other algorithms and conduct comparative studies. We aim to design more efficient threshold-raising strategies and explore novel pruning techniques to further reduce the search space. Finally, extending our work to advanced pattern types, such as top-$k$ periodic sequential rules, is also meaningful.

\section*{Data Availability}
The code and datasets are publicly available at \\ \href{https://github.com/DSI-Lab1/TPU}{https://github.com/DSI-Lab1/TPU.}

 \bibliographystyle{cas-model2-names}

\bibliography{paper}

\end{document}